\documentclass[final,3p,times,twocolumn,authoryear]{elsarticle}

\usepackage{amssymb}
\usepackage{amsmath}
\usepackage{graphicx}
\usepackage{xcolor}
\usepackage{booktabs}
\usepackage{supertabular}
\usepackage[version=4]{mhchem}
\usepackage{ulem}
\usepackage{threeparttable}
\usepackage{longtable}
\usepackage{caption}
\usepackage[
rightcaption
]{sidecap}
\usepackage{float}
\usepackage{placeins}
\usepackage{rotating}
\usepackage{txfonts}
\usepackage{hyperref}
\hypersetup{
    colorlinks=true,
    linkcolor= blue,
    filecolor=magenta,      
    urlcolor=blue,
    citecolor= blue,
}%

\journal{Life Sciences in Space Research}

\begin{document}

\begin{frontmatter}



\title{Overview of complex organic molecule observations in protostellar systems}


\author{Pooneh Nazari} 

\affiliation{organization={European Southern Observatory},
            addressline={Karl-Schwarzschild-Strasse 2}, 
            city={Garching},
            postcode={85748}, 
            country={Germany}}

\begin{abstract}
Complex organic molecules (COMs) have been detected abundantly at various stages of star formation, particularly in the warm protostellar phase. The progress in gas-phase measurements has been accelerated by the advent of the Atacama Large Millimeter/submillimeter Array and in ice measurements by the \textit{James Webb Space Telescope}. Particularly, the community has moved from single-source studies of COMs to statistical analyses because of these powerful instruments. In this article, I review surveys that consider COMs in the gas and ice. The two takeaways from this review include; 1. Gas-phase abundance ratios for some COMs show a small difference across many objects and the ice abundance ratios show similar or higher values to the gas, both pointing to the importance of ice chemistry in COM formation, 2. Some COM ratios show larger differences across many objects which could be due to either chemical or physical effects, thus both factors need to be considered when interpreting the data.

\end{abstract}



\begin{keyword}

Astrochemistry \sep Complex molecules \sep Protostellar systems \sep Submillimeter astronomy \sep Infrared astronomy 



\end{keyword}

\end{frontmatter}


\section{Introduction}
\label{sec:intro}

Complex organic molecules (COMs) are defined as species with at least six atoms that contain carbon. The formation and chemical evolution of these molecules have been the subject of many observational, modeling, and laboratory studies (see reviews by \citealt{Herbst2009}; \citealt{Caselli2012}; \citealt{Jorgensen2020}; \citealt{Ceccarelli2023}; \citealt{Jimenez2025}). Observationally, they are detected in multiple stages of star formation (\citealt{McGuire2018}), starting from the cold pre-stellar phase (\citealt{Bacmann2012}; \citealt{JimenezSerra2016}) to the warm protostellar phase (Class 0/I, \citealt{Ball1970}; \citealt{Dishoeck1995}), and more evolved protoplanetary disks (Class II, \citealt{Oberg2015}; \citealt{Walsh2016}). However, in the gas phase they are most easily detected in the protostellar phase, which is likely due to the higher temperatures of these systems (\citealt{vantHoff2020}; \citealt{Takakuwa2024}) and COM thermal sublimation into the gas. Thus in this paper, I only focus on COMs in protostellar systems. 

In ices, apart from methanol (CH$_3$OH; \citealt{Grim1991}), other COMs have only recently been detected and characterized in protostellar ices with the \textit{James Webb Space Telescope} (JWST, \citealt{Yang2022}; \citealt{Chen2024}; \citealt{Nazari2024_ices}; \citealt{Rocha2024}). However, hints of COMs in ices were found with the previous telescopes (\citealt{Schutte1999}; \citealt{Keane2001}; \citealt{Raunier2004}; \citealt{Terwisscha2018}; \citealt{Rachid2022}) and key simple species containing oxygen (O), nitrogen (N), and sulfur (S) were detected in the solid phase by \textit{Infrared Space Observatory} (ISO), Very Large Telescope (VLT), and \textit{Spitzer} (\citealt{Palumbo1995}; \citealt{Boogert1997, Boogert2008}; \citealt{Gibb2000, Gibb2004}; \citealt{Pontoppidan2003, Pontoppidan2008}; \citealt{Broekhuizen2005}; \citealt{Bottinelli2010}; \citealt{Oberg2011}). These species included CO, CO$_2$, H$_2$O, OCN$^-$, NH$_3$, and OCS (see \citealt{Boogert2015} review), many of which are thought to have an important role in formation of COMs in ices. 

\begin{SCfigure*}
    \includegraphics[width=0.77\textwidth]{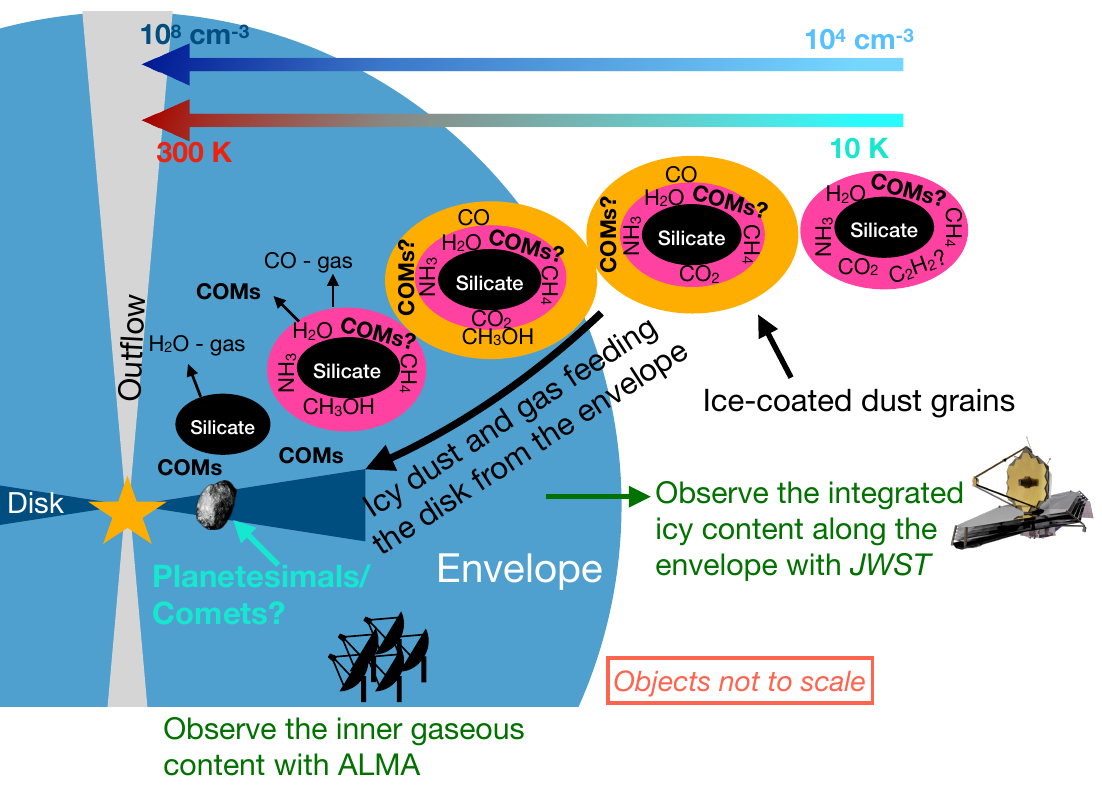}
    \caption{Cartoon showing different components of a protostellar system. The ices are thought to form in layers with different molecules at different temperatures and densities (see review by \citealt{Herbst2009}). Although JWST has detected COMs in ices, it is still under debate in which layer and at what stage of star formation they form. The icy dust grains and the gas in the envelope gets accreted onto the disk and eventually onto the planetesimals or comets that may be forming there.}
    \label{fig:cartoon}
\end{SCfigure*}

Study of COMs may also be relevant for planet habitability. Particularly, a remarkable correlation has been observed for many ice and gas COM abundances between the comet 67P and protostellar systems (\citealt{Drozdovskaya2019}; \citealt{Bianchi2019}; \citealt{Lippi2024}; \citealt{LopezGallifa2024}; \citealt{Rocha2024}). This likely points to at least a portion of cometary chemistry being set in the protostellar phase (Fig. \ref{fig:cartoon}). These comets may at a later stage deliver organics intact to planets (\citealt{Chyba1990}). Considering that some of the COMs could be precursors of more complex species necessary for habitable worlds (e.g., \citealt{Saladino2012}), understanding COM formation and evolution throughout the star formation process is crucial.

Laboratory and modeling studies suggest that COMs can form in the solid or the gas phase and the significance of these two scenarios is under debate (e.g., \citealt{Charnley1992}; \citealt{garrod2008complex, Garrod2022}; \citealt{Aikawa2008}; \citealt{Oberg2009}; \citealt{Balucani2015}; \citealt{Barone2015}; \citealt{Qasim2019}; \citealt{Enrique2025}). Laboratory studies of O-bearing COMs have found that many of them can form via hydrogenation of CO on the icy grains (\citealt{Fuchs2009}; \citealt{Fedoseev2015, Fedoseev2017}). Recently, another molecule, acetylene (C$_2$H$_2$) has been suggested as a starting point of COM formation before the CO freeze-out stage (Fig. \ref{fig:cartoon}; \citealt{Chuang2020, Chuang2021}), while C$_2$H$_2$ is yet to be detected in the prestellar phase. Formation of N-bearing COMs is less understood, likely due to the lower abundance of nitrogen in the interstellar medium (ISM) compared to oxygen (\citealt{Wilson1994}) and thus generally lower abundances of N-bearing COMs. Nevertheless, molecules such as OCN$^-$ and NH$_3$ with already secure detections in the ices (e.g., \citealt{dartois2001search}; \citealt{Broekhuizen2005}; \citealt{Bottinelli2010}) have been proposed to have an important role in chemistry of more complex N-bearing molecules such as HNCO and NH$_2$CH$_2$COOH (\citealt{Ligterink2018_HNCO}; \citealt{Ioppolo2021}). 

On the other hand, gas-phase reactions may still be at work for a select number of COMs in certain environments. For example, methyl cyanide (CH$_3$CN) shows efficient gas-phase formation routes in \cite{Garrod2022} models. Another molecule, formamide (NH$_2$CHO), might also have gas-phase formation routes based on quantum computations (\citealt{Barone2015}). Acetaldehyde (CH$_3$CHO) was suggested to have two efficient gas-phase formation routes by \cite{Vazart2020}. It is worth noting that the chemistry of these three molecules is under active debate and both solid and gas-phase formation routes have been suggested for them (\citealt{Huntress1979}; \citealt{Walsh2014}; \citealt{Skouteris2017}; \citealt{Dulieu2019}; \citealt{Fedoseev2022}).

To better understand the formation of COMs, observers have searched for them in a variety of environments in the ISM. They have measured their abundances in the gas and ices in a quest to answer questions such as \textit{how universal is their chemistry?} or \textit{how similar are their abundances across different environments?} In the past few years, the community has moved from single-object studies to large-sample statistical analysis of COMs, particularly in the gas-phase. This is thanks to the unprecedented sensitivity of modern telescopes such as the Atacama Large Millimeter/submillimeter Array (ALMA) and JWST, and their ability to produce deep observations of large number of sources in a relatively short period of time. In this work, I review the state of the art of surveys that consider COMs in the gas and ice, their major findings, and a possible path forward.   

Section \ref{sec:teles_tech} briefly describes the techniques used to detect and quantify COMs in the gas and ice. Section \ref{sec:surveys} outlines the recent gas and ice surveys considering COMs. Sections \ref{sec:takeaway} and \ref{sec:origin} discuss the key takeaways of the surveys and possible origins of several observational discrepancies. Finally, in Sect. \ref{sec:conclusion}, I conclude and propose a path for the future.

\section{Telescopes and techniques}
\label{sec:teles_tech}
\subsection{Gas phase}

\begin{figure*}
    \centering
    \includegraphics[width=0.95\textwidth]{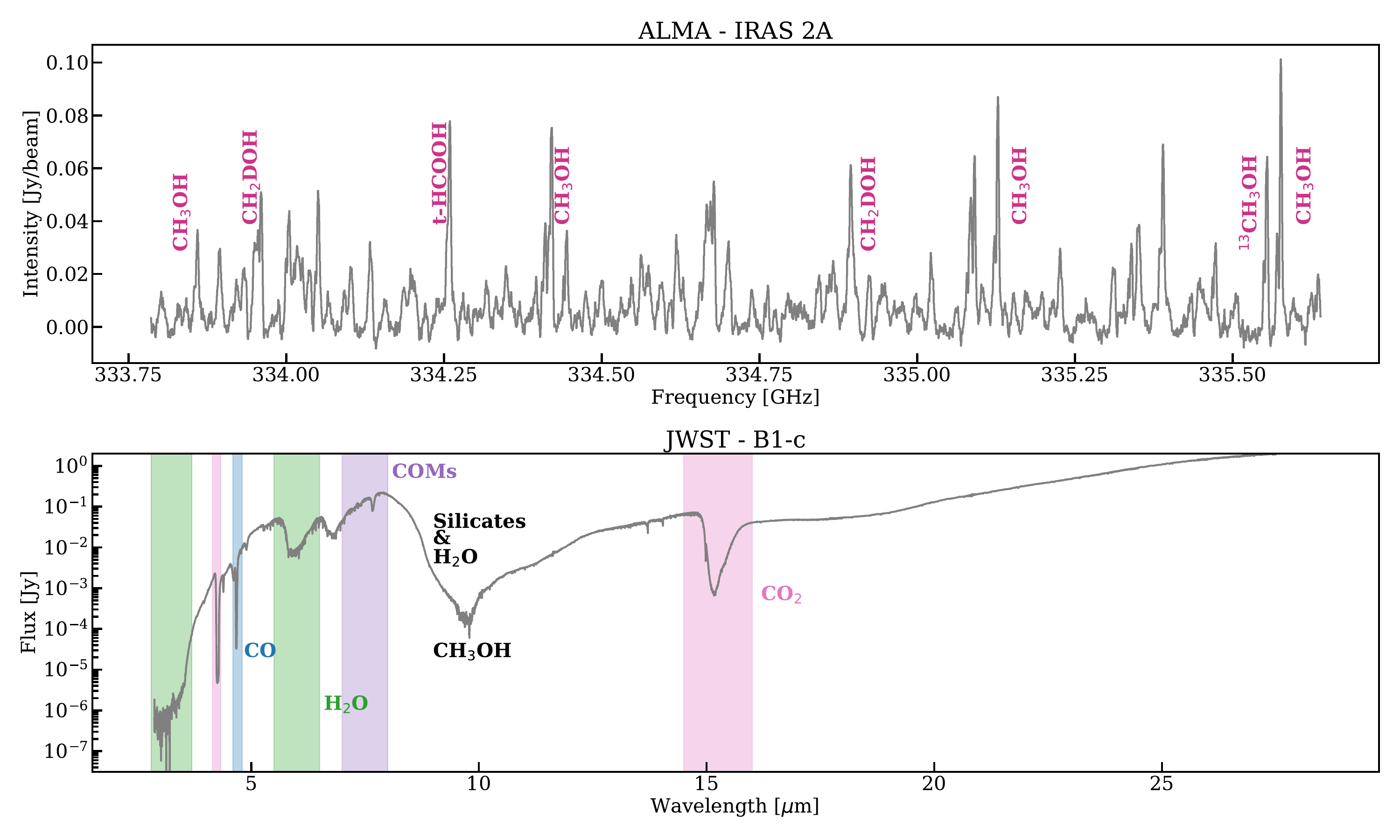}
    \caption{Example spectra of ALMA (top) and JWST (bottom) to showcase the typical molecular observations in the gas and ice. The ALMA data are taken from \cite{Chen2024} and extracted at the peak of the continuum. The JWST data are taken from the JOYS+ program (Sect. \ref{sec:JOYS}) and the spectrum is extracted from a cone aperture centered on the continuum at 5.25\,$\mu$m.}
    \label{fig:spec}
\end{figure*}

Molecules in the gas phase can be detected via their rotational and vibrational transitions. These molecules can either emit or absorb depending on the gas temperature. Complex organics have been detected in the gas phase at a range of wavelengths (e.g., \citealt{wang2011herschel}; \citealt{neill2012laboratory}; \citealt{Neill2014}) but here I focus on millimeter and submillimeter observations (top panel of Fig. \ref{fig:spec}). At these wavelengths, COMs were first detected with single dish telescopes (\citealt{Sutton1985}; \citealt{Blake1987}; \citealt{Dishoeck1995}; \citealt{Gibb2000_hot}; \citealt{Cazaux2003}) and later by interferometers such as Submillimeter Array (SMA), Northern Extended Millimeter Array (NOEMA), and ALMA with better angular resolution and sensitivity (see review by \citealt{Jorgensen2020}). Although single-dish studies of these molecules are still relevant, here I focus on interferometric studies. Among the modern interferometers, ALMA particularly revolutionized the study of COMs by better revealing their spatial distribution and measuring their column densities with as high a level of accuracy as ${\sim}20-30\%$. Therefore, I discuss a few recent surveys from the SMA and NOEMA but mostly focus on the ALMA studies.     

Often when studying COMs in the gas phase at millimeter and submillimeter it is assumed that these molecules are in local thermodynamical equilibrium (LTE). This assumes that the densities are high enough for the excitation and de-excitation of COMs being dominated by collisions. Therefore, all transitions of the molecule can be explained by Boltzmann distribution at a single temperature. That being said, a few recent works have noted the potential importance of non-LTE effects in column density measurements of some COMs (e.g., \citealt{Bianchi2020}; \citealt{Frediani2025}). Measurement of COM column densities and excitation temperatures can be achieved using spectral analysis tools such as Weeds, CASSIS, XCLASS, and MADCUBA  (\citealt{Maret2011}; \citealt{Vastel2015}; \citealt{Moller2017}; \citealt{Martin2019}). These measurements use COM rotational spectroscopy from laboratory works (see review by \citealt{widicus2019millimeterwave}) and often their compilation in major databases (\citealt{Pickett1998}; \citealt{Muller2001, muller2005}). Finally, there have been some recent developments analyzing line-rich spectra using machine learning techniques (e.g., \citealt{lee2021machine}; \citealt{fried2023implementation}). These methods particularly speed up the process at the line identification stage before detailed analysis of the spectra and even predict the most likely species that could exist within a system but not yet detected.

\subsection{Solid phase}
\label{sec:solids}

In the solid phase, molecules can only be detected via their vibrational transitions. Although it is possible to detect them in emission (\citealt{Malfait1999}; \citealt{Molinari1999}), normally they get sublimated before being sufficiently thermally excited to emit. Therefore, they are mostly detected in absorption (bottom panel of Fig. \ref{fig:spec}). Detection of ices in absorption has been facilitated by infrared observatories such as the ISO, VLT, \textit{Spitzer}, and now JWST. In terms of COMs, only methanol, the most abundant and common COM, was securely detected in the pre-JWST era (e.g., \citealt{Grim1991}; \citealt{Keane2001}; \citealt{Gibb2004}; \citealt{Bottinelli2010}). However, JWST revolutionized this field by detecting other COMs for the first time in ices (e.g., \citealt{Rocha2024}). Here, I mainly focus on JWST surveys of COMs.     

Column densities of molecules in ices are measured after converting the spectral flux into optical depth (see review by \citealt{Boogert2015}). The integration over the ice feature in optical depth space can then be turned into column density using the band strength ($A$). Although JWST has detected these molecules in ices, major uncertainties remain. One important issue in ice analysis is the large width of the ice absorption features. This results in a single observed ice feature being a superposition of multiple bands. Other major uncertainties include the continuum determination and the uncertainty on the band strength which is normally on the order of ${\sim}30\%$ (e.g., \citealt{Terwisscha2018}; \citealt{Rachid2022}). A complexity with ices is that the shape of the absorption feature (i.e., its width and peak position) for a molecule depends on its ice matrix. However, this is also an opportunity to better understand the ice environment in which a molecule resides. Some of these complications can be mitigated by using methods such as those performed by the \texttt{ENIIGMA} fitting tool (\citealt{Rocha2021}), where the significance of a particular molecule's infrared spectrum is examined statistically. Finally, it is worth noting that the current COM studies with JWST are only possible because of years of accumulated laboratory work that provide the spectroscopy of various COMs and their compilation in databases (see \citealt{Rocha2022} and the references therein). However, this is still a limitation because regardless of these huge efforts, laboratory spectra do not exist for all molecules and in all ice mixtures. As a result of all the above, final uncertainty on column density of each molecule boils down to multiple factors and is different per molecule. It can be as small as ${\sim}30\%$ or as large as ${\sim}70\%$.

\section{Surveys of COMs in protostellar systems}
\label{sec:surveys}

\subsection{Gas phase}

In this section, I give an overview of the current gas-phase surveys considering COMs in the literature. Table \ref{tab:gas} presents a summary of the gas-phase surveys. This section is mainly aimed to present the main results of these surveys rather than an exhaustive report. It is worth noting that most of the studies discussed here have medium angular resolution of ${\sim}0.3''-1''$. Therefore, the exact regions where these COMs trace are still debated. Another important bias in the current surveys is that not all are sensitive enough to detect the less abundant COMs. This problem is even more severe for low-mass objects where due to their lower luminosities, masses, and temperatures, higher sensitivities are required to observe the less abundant molecules. Therefore, the current measurements of less abundant COMs are dominated by high-mass sources and our statistical understanding of formation and evolution of these less abundant COMs might be biased by those higher mass objects.

\begin{sidewaystable*}
\small
\begin{threeparttable}
    \caption{Gas-phase surveys}
    \label{tab:gas}
    \centering
    \setlength{\tabcolsep}{4pt}
    \begin{tabular}{l l l l l l l l} 
    \toprule
    \toprule 
Survey                         & Telescope   &   Ang. res. [$''$]               &  Wavelength [mm]      &     No. of objects    &    Type of objects$^*$    &  COM detection rate$^\dagger$               &  N/O chemical differentiation$^\star$ \\
\midrule  
\hyperref[sec:ALCHEMI]{ALCHEMI}       & ALMA        &      ${\sim}1.6$                 &   ${\sim}0.8-3.5$     &       1               &    Starburst galaxy   &        ---                        &               Deficiency of N-bearing COMs in one giant              \\
&&&&&&&molecular cloud\\
\hyperref[sec:ALMAGAL]{ALMAGAL}       & ALMA        &      ${\sim}0.2$                 &   ${\sim}1.4$         &       ${\sim}1000^{\rm d}$    &    Mainly high-mass   &  CH$_3$OH, ${\sim}60\%^{\rm a}$   &  On average, CH$_3$OH tracing the outer layers  \\
&&&&&&& while CH$_3$CN and CH$_3$OCHO trace the \\
&&&&&&& dense dust fragments.\\
\hyperref[sec:ALMA-IMF]{ALMA-IMF}      & ALMA        &      ${\sim}0.3-1.5$             &   ${\sim}1.3,\, 3$    &       ${\sim}800$     &    Low and high-mass  &  CH$_3$OCHO, ${\sim}10\%$         &  O-bearing molecules traced the outflow or shocks\\
&&&&&&&toward at least two objects\\
\hyperref[sec:ALMASOP]{ALMASOP}       & ALMA        &      ${\sim}0.35,\, 1,\, 7$      &   ${\sim}1.3$         &       56              &    Class 0/I    &  CH$_3$OH, $20-40\%^{\rm b}$      &  ---\\
\hyperref[sec:ATOMS]{ATOMS}         & ALMA        &      ${\sim}1-2$                 &   ${\sim}3$           &       453             &    Mainly high-mass   &  COMs, $20-30\%^{\rm c}$          &  Spatial differentiation in emission peaks of \\
&&&&&&& C$_2$H$_5$CN and CH$_3$OCHO in ${\sim}50\%$ of hot cores.\\
\hyperref[sec:ATOMS]{QUARKS}        & ALMA        &      ${\sim}0.3$                 &   ${\sim}1.3$         &       1562             &    Mainly high-mass   &        CH$_3$CONH$_2$, ${\sim}20\%$                         &           ---\\
\hyperref[sec:CALYPSO]{CALYPSO}       & PdBI       &      ${\sim}0.4-1$               &   ${\sim}1.3, 3$      &       26              &    Class 0/I          &  CH$_3$OH, ${\sim}50\%$           &   Objects associated with either O-bearing, \\
&&&&&&& CHO-bearing, or cyanide groups.\\
\hyperref[sec:CoCCoA]{CoCCoA}        & ALMA        &      ${\sim}0.3$                 &   ${\sim}1.2$         &       23$^{\rm d}$    &    High-mass          &        ---                        &       ---\\
\hyperref[sec:DIHCA]{DIHCA}         & ALMA        &      ${\sim}0.3$                 &   ${\sim}1.3$         &       30$^{\rm d}$    &    Low and high-mass  & $^{13}$CH$_3$OH, ${\sim}40\%$     &   Spatial distribution of NH$_2$CHO is more compact \\
&&&&&&&than HNCO and CH$_3$CN likely related to \\
&&&&&&& their sublimation temperatures.\\
\hyperref[sec:EMoCA]{EMoCA}         & ALMA        &      ${\sim}1-2$                 &   ${\sim}3$           &       1$^{\rm d}$     &    High-mass          &        ---                        &       ---\\
\hyperref[sec:EMoCA]{ReMoCA}        & ALMA        &      ${\sim}0.3-0.9$             &   ${\sim}3$           &       1$^{\rm d}$     &    High-mass          &        ---                        &   Spatial differentiation of N- vs O-bearing\\ 
&&&&&&& COMs toward SgrB2(N1).\\
\hyperref[sec:FAUST]{FAUST}         & ALMA        &      ${\sim}0.25-0.35$           &   ${\sim}1.3,\, 3.5$  &       13$^{\rm d}$             &    Class 0/I          &       ---        &   COMs tracing an onion-like structure in IRAS 4A2.\\
\hyperref[sec:GUAPOS]{GUAPOS}        & ALMA        &      ${\sim}1.2$                 &   ${\sim}3$           &       1$^{\rm d}$     &    High-mass          &       ---                         &   CH$_3$CHO and CH$_3$OHCHO vs. other O- and\\
&&&&&&& N-bearing COMs (differentiation in line\\
&&&&&&& velocities and peak positions).\\
\hyperref[sec:PILS]{PILS}          & ALMA        &      ${\sim}0.5$                 &   ${\sim}0.8$         &       1$^{\rm e}$     &    Class 0            &       ---                         &   COMs tracing an onion-like structure toward \\ 
&&&&&&& IRAS16293 A consistent with their\\
&&&&&&& sublimation temperatures. \\
\hyperref[sec:PILS-Cygnus]{PILS-Cygnus}   & SMA         &      ${\sim}1$                   &   ${\sim}0.8$         &       10              &    Intermediate/High-mass &   $^{13}$CH$_3$OH, 40\%       &   Spatial differentiation of O-bearing vs. \\
&&&&&&& N- and S-bearing species in one system. \\
\hyperref[sec:PEACHES]{PEACHES}       & ALMA        &      ${\sim}0.5$                 &   ${\sim}1.2$         &       50              &    Class 0/I          &  CH$_3$OH, 56\%                   &   ---\\
\hyperref[sec:PRODIGE]{PRODIGE}       & NOEMA       &      ${\sim}1$                   &   ${\sim}1.3$         &       30              &    Class 0/I          &   $^{13}$CH$_3$OH, ${\sim}23\%$                             &   ---\\
\hyperref[sec:PEACHES]{ORANGES}       & ALMA        &      ${\sim}0.25$                &   ${\sim}1.3$         &       28              &    Class 0/I          &   CH$_3$OH, 26\%                  &   ---\\
\hyperref[sec:SOLIS]{SOLIS}         & NOEMA       &      ${\sim}2-4$                 & ${\sim}1.5,\, 3,\, 4$ &       7$^{\rm f}$               &    A range$^{\rm f}$  &    ---                            &  Potential chemical differentiation \\
&&&&&&& along the outflows of IRAS 4A system \\
&&&&&&& and in the shocked region L1157-B1.\\

\bottomrule
\end{tabular}

\begin{tablenotes}
        \item[$*$]: Number of continuum sources, unless otherwise specified.
        \item[$\dagger$]: The molecule used for the measurement and the detection rate.
        \item[$\star$]: A few examples of nitrogen-oxygen differentiation observed in the samples.
\item[a]: The detection rate based on CH$_3$CN and CH$_3$OCHO is ${\sim}35\%$ and ${\sim}11\%$.
\item[b]: The range considers different criteria.
\item[c]: Based on detection of at least 5 COM lines without considering one specific COM.
\item[d]: Star forming region.
\item[e]: System of multiple low-mass protostars.
\item[f]: Objects include Class 0/I objects, a prestellar core, a protocluster, and a shock region.
    \end{tablenotes}
\end{threeparttable}
\end{sidewaystable*}

\subsubsection{ALCHEMI}
\label{sec:ALCHEMI}

The ALMA Comprehensive High-resolution Extragalactic Molecular Inventory (ALCHEMI; PI: S. Mart{\'\i}n; \citealt{Martin2021}) program observed the starburst galaxy NGC253 central molecular zone with Bands 3-7 of ALMA. The current studies from this program mainly focus on simple molecules (e.g., \citealt{Harada2021}; \citealt{Holdship2022}; \citealt{Gong2025}), however, using their subset ACA data \cite{Martin2021} reported the first detection of complex species such as C$_2$H$_5$OH and HC$_3$HO in an extragalactic ISM. They also found that the HCOOH/CH$_3$OH and C$_2$H$_5$OH/CH$_3$OH ratios from NGC253 is in agreement with and on the higher end of those from the Galactic hot cores and Galactic Center giant molecular clouds. Another study considered CH$_3$CHO, C$_2$H$_5$OH, NH$_2$CHO, CH$_2$NH, and CH$_3$NH$_2$, finding that these molecules generally traced the inner regions of the central molecular zone (\citealt{Bouvier2025}). They also found different gas components and origins for the emission from these COMs, where the COMs could be affected by shock and/or thermal processes. Finally, they found that one of the giant molecular clouds in the region showed deficiency in N-bearing COMs. They interpreted this as this cloud either having a different evolutionary stage or simply having lower level of star formation activity.

\subsubsection{ALMAGAL}
\label{sec:ALMAGAL}
ALMA Evolutionary study of High Mass Protocluster Formation in the Galaxy (ALMAGAL; PI: Sergio Molinari, \citealt{Molinari2025}) survey covered ${\sim}1000$ dense clumps, with not all being line rich. The current ALMAGAL studies mainly focus on other aspects of star formation such as fragmentation or filamentary structures (\citealt{Wells2024}; \citealt{Coletta2025}; \citealt{Sanchez2025}). However, a recent work from this team considered detection statistics of various species including a few COMs in the sample (\citealt{Mininni2025}). Interestingly they found that the detection rate of methanol in their large sample of ${\sim}1000$ clumps at different evolutionary stages was on the order of 60\%. This number for methyl cyanide and methyl formate (CH$_3$OCHO) was ${\sim}35\%$ and ${\sim}11\%$, respectively. They also found that CH$_3$OH emission does not follow the continuum while those of CH$_3$OCHO and CH$_3$CN follow the dense dust fragments. Other studies have also used this rich dataset to analyze methanol (\citealt{vanGelder2022}), a few N-bearing molecules (\citealt{Nazari2022_ALMAGAL}; \citealt{Nazari2023}), and methanol deuteration (\citealt{vanGelder2022_deuteration}) in a sub-sample of their objects. 

In a study combining the data from PEACHES (Sect. \ref{sec:PEACHES}) and ALMAGAL programs, \cite{vanGelder2022} measured the methanol column densities toward 148 low- and high-mass objects. They primarily used methanol minor isotopologues to find the column density of the major isotopologue. They found a large spread (${\sim}4$ orders of magnitude) in the warm methanol mass for objects with similar luminosities. Comparing their results with a simple spherical toy model, they concluded that the temperature structure of the objects with large warm methanol mass is likely less affected by presence of a disk. However, objects with large disk radii (${\gtrsim}50$\,au) showed up to two orders of magnitude lower warm methanol mass. This led them to conclude that the temperature structure of those objects could have been significantly affected by the disks. They also found that optically thick dust can hide the emission (also see \citealt{DeSimone2020}).

In the sample studied by \cite{Nazari2022_ALMAGAL}, CH$_3$CN, HNCO, C$_2$H$_5$CN, C$_2$H$_3$CN, and NH$_2$CHO were detected in more than 30 high-luminosity protostellar systems. They found that the column density ratios of most of their N-bearing COMs were relatively constant across systems with a wide range of luminosities and masses, pointing to formation of those COMs in similar physical conditions, likely prestellar ices. The only exception was NH$_2$CHO which showed larger variations across systems. Based on the high average excitation temperature of this molecule, they speculated that a reason for these variations could be physical rather than chemical. Moreover, \cite{Nazari2023} used the same sample (${\sim}40$ protostellar systems) and found that COMs not containing oxygen systematically showed higher abundances for their hot component in regions closer to the protostar ($T{\gtrsim} 300$\,K) compared to their warm component in regions further away ($T{\sim}150$\,K). However, they did not find this enhancement for molecules containing oxygen. This led them to conclude that this enhancement might be due to destruction of carbon grains closer to the protostars (also see \citealt{vantHoff2020_carbon}; \citealt{Walls2024}; and \citealt{Law2025}).

In the methanol deuteration study, \cite{vanGelder2022_deuteration} found that D/H ratios from CH$_2$DOH/CH$_3$OH were systematically lower in the high-mass objects of an ALMAGAL sub-sample compared with the lower-mass objects of the literature. By comparison with models, they concluded that this is either due to higher temperatures of the prestellar phase or shorter prestellar timescales for high-mass objects. Another interesting finding of their work was that D/H found from CHD$_2$OH/CH$_2$DOH ratios were similar between low- and high-luminosity systems and all were at least one order of magnitude higher than those found from CH$_2$DOH/CH$_3$OH. Based on their average CHD$_2$OH/CH$_2$DOH, they concluded that ${\sim}1/5$ of singly deuterated methanol molecules get deuterated further to form CHD$_2$OH in low- and high-mass systems.

\subsubsection{ALMA-IMF}
\label{sec:ALMA-IMF}
The ALMA-IMF large program (PI: F. Motte; \citealt{Motte2022}) observed 15 massive clouds and its main objective is to measure the core mass function which can be compared with the initial mass function (IMF) of stars (e.g., \citealt{Pouteau2022, Pouteau2023}; \citealt{Louvet2024}). However, a few studies looked at the chemistry in their sample. For example, one study from this program considered CH$_3$CN, CH$_3$OCHO, and CH$_3$CCH toward a massive star forming region (\citealt{Brouillet2022}; also see \citealt{Molet2019}). In this region, they found seven hot cores with masses between 16 and 100\,M$_{\odot}$ and one with lower mass of about 2\,M$_{\odot}$. The normalized line intensity ratios among their objects with orders of magnitude differences in mass were similar to within a factor of ${\sim}2-3$. \cite{Brouillet2022} concluded that this likely points to these sources having similar chemical compositions. They also found similar excitation temperatures for CH$_3$CN toward the hot cores (120\,K-160\,K), while the temperatures found from CH$_3$CCH were lower (50\,K-90\,K). They concluded that this is likely due to CH$_3$CCH tracing the more extended envelope. For one of their objects, the emission from a few O-bearing molecules was found to be likely associated with the outflow lobes. Thus those molecules may have been released from the ices through shocks or UV irradiation in the outflow cavities.

The chemistry in the larger sample was studied by \cite{Bonfand2024} who identified 76 compact methyl formate objects (masses between ${\sim}0.2$\,M$_{\odot}$ and ${\sim}80$\,M$_{\odot}$) with about 30\%-50\% of them having core masses larger than 8\,M$_{\odot}$ (for comparison between luminosities found from COM lines of \citealt{Bonfand2024} and dust emission see \citealt{Motte2025}). Another study from this team found the same hot cores as \cite{Bonfand2024} toward one star-forming region when considering detection of other COM lines in their analysis (\citealt{Armante2024}). \cite{Bonfand2024} also found that the methyl formate emission toward lower mass objects may be explained by either shocks or associated with more evolved objects. It is also worth noting that from the 807 compact continuum sources analyzed in \cite{Bonfand2024}, only 76 showed methyl formate which is ${\sim}9\%$ of their sample and very much inline with the results of \cite{Mininni2025} for the ALMAGAL sample. Finally, they considered the relation between number of methyl formate objects and the mass, evolutionary stage, and the number of continuum sources per protocluster. Even though they did not find any relation between number of methyl formate objects and mass or evolutionary stage, suggesting that emergence of hot cores is independent of the global protocluster properties, they did find a positive correlation between number of methyl formate objects and the number of continuum sources. Finally, it is worth noting that the ALMA-IMF setup includes transitions from multiple COMs (\citealt{Motte2022}) and further chemical studies of their hot cores will likely follow in future publications.

\subsubsection{ALMASOP}
\label{sec:ALMASOP}
The ALMA Survey of Orion Planck Galactic Cold Clumps (ALMASOP; PI: T. Liu) observed 72 clumps in the Orion complex where not all are line rich. From those \cite{Dutta2020} identified 56 Class 0/I protostellar systems. Many of the current works from this program focus on jets and outflows, multiplicity, or physical structure of the clumps (e.g., \citealt{Luo2022}; \citealt{Jhan2022}; \citealt{Hirano2024}; \citealt{Liu2025}). From their COM studies, \cite{Hsu2022} found 11 objects out of the 56 that showed at least two detected methanol lines. Using the same criteria as \cite{Yang2021}, they found that statistically, the detection rates of warm methanol were similar between the Perseus ALMA Chemistry Survey (PEACHES) (Sect. \ref{sec:PEACHES}) and ALMASOP. They also considered the correlation between CH$_3$CHO, CH$_3$OCHO, C$_2$H$_5$OH, and CH$_2$DOH with CH$_3$OH, finding that the ratios of CH$_3$CHO/CH$_3$OH being more diverse than those of CH$_3$OCHO/CH$_3$OH. They concluded that this agrees with the results of previous work on these molecules (\citealt{vanGelder2020}) and their suggestion that likely the less diverse ratios point to formation in cold early stages, while the higher diversity may point to local source properties. \cite{Hsu2022} also found a correlation between CH$_3$CN and CH$_3$OH similar to the results of PEACHES (\citealt{Yang2021}; Sect. \ref{sec:PEACHES}) and Continuum And Lines in Young ProtoStellar Objects (CALYPSO; \citealt{Belloche2020}; Sect. \ref{sec:CALYPSO}) surveys regardless of an obvious chemical link between the two molecules.

Another work used the Atacama Compact Array (ACA) of ALMA to study four low-mass protostellar systems (\citealt{Hsu2020}). They detected multiple COMs toward their objects with excitation temperatures and line widths suggesting that they trace the inner warm regions. In one of their objects they also found that the D/H and $^{12}$C/$^{13}$C from methanol was comparable to other low-mass protostellar systems. A more recent study from this survey analyzed one of the ALMASOP objects in more detail and found a rotating structure rich in COMs (\citealt{Hsu2025}). They concluded that these COMs are likely tracing the Keplerian disk or near its boundary. Accretion shocks were proposed as a plausible scenario to explain the origin of the emission.

\subsubsection{ATOMS and QUARKS}
\label{sec:ATOMS}

The ALMA Three-millimeter Observations of Massive Star-forming regions (ATOMS; PI: T. Liu) surveyed 146 star-forming regions with Band 3 at an angular resolution of ${\sim}1''-2''$. Their main focus is on better understanding of star formation and the role of feedback and filaments in this process (e.g., \citealt{Liu2020}; \citealt{Liu2022_Hong}; \citealt{Saha2022}). \cite{Liu2021} detected 453 sources in the 3\,mm continuum with 32 showing COM-rich spectra (${\geq}20$ lines) and 58 showing a lower level of COM-richness. These two categories were not associated with hyper- or ultra-compact H$_{\rm II}$ regions. Including those associated with hyper- or ultra-compact H$_{\rm II}$ regions, each group will be around 24 objects larger, resulting in total of 138 objects with COM signatures.

\cite{Qin2022} conducted a more focused study of the line-rich spectra. They found chemical differentiation among O- and N-bearing molecules in 29 of their 60 objects and a tight correlation between CH$_3$OCHO and CH$_3$OH. Another chemical study focused on C$_2$H$_5$OH and CH$_3$OCH$_3$ in the sample of 60 objects (\citealt{Kou2025}) that 39 of them showed simultaneous detection of these two molecules. They found strong correlations between C$_2$H$_5$OH, CH$_3$OCH$_3$, and CH$_3$OH column densities. Moreover, they found that the ratio of C$_2$H$_5$OH/CH$_3$OCH$_3$ is constant. They concluded that these point to likely chemical connection of these molecules, with CH$_3$OH a potential precursor.

Another study from this team considered one star forming region (\citealt{Peng2022}) and found two cores in that region with chemical differentiation. One being rich in O-bearing COMs and the other rich in N-bearing molecules and particularly HC$_3$N. They associated this differentiation in N/O to likely difference in their initial temperatures at the accretion phase. They also found evidence for CH$_3$OCHO and CH$_3$OCH$_3$ being chemically linked, in addition to C$_2$H$_5$OH and CH$_3$OCH$_3$ having methanol as a common precursor.

The Querying Underlying mechanisms of massive star formation with ALMA-Resolved gas Kinematics and Structures (QUARKS; PI: L. Zhu; \citealt{Liu2024}) observed 139 massive clumps from the ATOMS survey. They used higher angular resolution (${\sim}0.3''$) with ALMA Band 6 to resolve features previously unresolved by ATOMS (e.g., \citealt{Yang2024}). They detected 207 1.3\,mm continuum sources from ALMA's 7\,m compact array (\citealt{Xu2024}) and 1562 cores when including the data from the 12\,m array (\citealt{Yang2025}). One study already considered the chemistry in a portion of their objects (\citealt{Duan2025}). They searched for acetamide (CH$_3$CONH$_2$) toward 52 hot cores while detecting it in 10 of those (detection rate of ${\sim} 20\%$). They found a relatively constant ratio for NH$_2$CHO/CH$_3$CONH$_2$ toward their objects and interpreted this as a sign of chemical link between these molecules likely through solid-phase reactions. Their spectral setup covers transitions from various other COMs (\citealt{Liu2024}), with more chemical studies likely a focus of future studies.

\subsubsection{CALYPSO}
\label{sec:CALYPSO}

The CALYPSO survey (PI: P. Andr{\'e}) is a large program with Plateau de Bure Interferometer (PdBI). It covers 26 Class 0/I objects with an angular resolution of ${\sim}0.5''$. Although this survey has studied other aspects of star formation (e.g., \citealt{Maret2014}; \citealt{Maury2019}; \citealt{Podio2021}), the major COM analysis for this sample was done by \cite{Belloche2020}. This work found that not all the objects in their sample show COM emission. These objects were those with luminosities $< 2$\,L$_{\odot}$ and thus the non-detection might be related to lack of sensitivity. One of the main conclusions of \cite{Belloche2020} was that the COM emission could have different origins; with the most likely scenario being the canonical inner envelope/disk origin but also some of the COM emission might be related to shocks and outflows. They also identified three groups of objects based on the abundances of CN-bearing, O-bearing, and CHO-bearing molecules with respect to methanol. They concluded that these variations among systems might be related to evolutionary or local environmental effects. Finally, an important result of \cite{Belloche2020} was that they found correlations between molecules that did not have a known chemical link. Therefore, they concluded that observing a correlation among two molecules does not necessarily imply a chemical link between the two.    

\subsubsection{CoCCoA}
\label{sec:CoCCoA}

Complex Chemistry in hot Cores with ALMA (CoCCoA; PI: B. McGuire) observed 23 high-mass star forming regions at ${\sim}0.3''$. Currently, there are two studies out from this program on COMs. One which considers various O-bearing COMs (\citealt{Chen2023}) and the other that focuses on acetone (CH$_3$COCH$_3$) in the sample and compares that with ice abundances (\citealt{Chen2025}). The former, studied CH$_3$CHO, C$_2$H$_5$OH, CH$_3$OCH$_3$, CH$_3$OCHO, CH$_2$OHCHO, and (CH$_2$OH)$_2$ toward 14 high-mass systems. Generally, they found similar column density ratios with respect to methanol for low- and high-mass systems, with CH$_3$OCH$_3$/CH$_3$OH and CH$_3$OCHO/CH$_3$OH showing a smaller scatter than the others. They concluded that the constant ratios among low- and high-mass objects points to them forming in similar physical environments and likely in the prestellar ices (also see \citealt{Coletta2020}), while the larger scatter may be due to chemical or physical factors. The other study focused on CH$_3$COCH$_3$, CH$_2$CO, CH$_3$CCH, and C$_2$H$_5$CHO in 12 of the CoCCoA objects (\citealt{Chen2025}). They detected the first three molecules in their 12 objects, while C$_2$H$_5$CHO was only tentatively detected. They found that CH$_3$COCH$_3$, CH$_2$CO, and C$_2$H$_5$CHO likely have a hot core origin, while CH$_3$CCH could have an outflow origin. Comparing the gas and ice abundances of acetone with respect to methanol, they concluded that the higher ice abundances (by a factor of ${\sim}10$) may point to further gas-phase reprocessing of acetone after sublimation.        

\subsubsection{DIHCA}
\label{sec:DIHCA}

Digging into the Interior of Hot Cores with ALMA (DIHCA; PI: P.\ Sanhueza) survey observed 30 high-mass star-forming regions. Multiple works from this program consider the massive star formation process and clump fragmentation (\citealt{Olguin2021}; \citealt{Olguin2022}; \citealt{Ishihara2024}). However, two studies from this team focused on chemistry. \cite{Taniguchi2023} studied NH$_2$CHO, HNCO, H$_2$CO, and CH$_3$CN toward the 30 regions. They found strong correlations between NH$_2$CHO and the two molecules HNCO and H$_2$CO. They concluded that this points to their chemical link forming in the gas, where chemical models including those gas-phase reactions can explain the observed abundances. Another study considered deuterium fractionation of methanol in the higher-mass cores ($> 10$\,M$_{\odot}$) of the DIHCA sample (\citealt{Sakai2025}), where they found $^{13}$CH$_3$OH toward ${\sim}40\%$ of those cores. They found a lower CH$_2$DOH/CH$_3$OH in their high-mass star forming regions compared with those of low-mass star-forming regions in the literature. Using chemical models they found that objects with shorter cold phase would have lower CH$_2$DOH/CH$_3$OH. They concluded that the lower abundances toward high-mass star-forming regions is likely due to this shorter cold phase and the diversity in the CH$_2$DOH/CH$_3$OH ratio among high-mass objects may also be explained by the diversity in the timescale of the prestellar phase.  

\subsubsection{EMoCA and ReMoCA}
\label{sec:EMoCA}

The Sagittarius B2 (SgrB2) region has been observed by two surveys with a large frequency bandwidth (${\sim}30$\,GHz). These two are called Exploring Molecular Complexity with ALMA (EMoCA; PI: A. Belloche) and Re-exploring Molecular Complexity with ALMA (ReMoCA; PI: A. Belloche) surveys, where the latter has higher angular resolution and sensitivity. This team provided multiple first molecular detections (\citealt{Belloche2014}; \citealt{Belloche2019}) and detailed chemical models to explain their abundances (\citealt{Garrod2017}; \citealt{Willis2020}). One interesting finding of this team was that the deuteration level of multiple COMs were lower than prediction from chemical models and abundances found toward Orion KL (\citealt{Belloche2016}). They concluded that this could be either due to high temperatures of the Galactic Center region or generally lower deuterium abundances in those regions. Another work considered CH$_3$OH, C$_2$H$_5$OH, CH$_3$SH, and C$_2$H$_5$SH in this region (\citealt{Muller2016}). They firmly detected the first three and found that ratios among those molecules are consistent between SgrB2(N2) and Orion KL. Using the 3\,mm data of EMoCA, \cite{Bonfand2017} found three additional cores toward SgrB2 with similar chemical compositions but differing from that of SgrB2(N2). A later study using the same dataset and chemical models showed that the COM abundances with respect to methanol toward SgrB2(N2-N5) are well-explained with cosmic ray ionization rate of ${\sim}7 \times 10^{-16}$\,s$^-1$ (\citealt{Bonfand2019}). They also found that COMs could efficiently form on dust grains with minimum temperatures of 15\,K in the pre-stellar phase, while those of 25\,K are too high.

Using the 3\,mm portion of the ReMoCA program, \cite{Busch2022} analyzed COM distribution toward the N1 source. They found that from the analyzed COMs, the N-bearing ones peaked at higher temperatures, while most O-bearing ones peaked at lower temperatures. They also found that COMs which mainly form on grains desorb together with water at around 100\,K, without any dependence on their binding energy. However, they also found COM emission at lower temperatures which might be explained by non-thermal desorption mechanisms or lower binding energies in the outer ice layers on the grains which are expected to be water-poor. A more recent study (\citealt{Busch2024}) used the same dataset to compare COM abundances in a region affected by the outflow and one that is not toward SgrB2(N1). They found that the N-bearing molecules were enhanced toward the outflow position. They concluded that due to an outflow-driven shock wave many O-bearing species got destroyed while the N-bearing molecules such as cyanides and cyanopolyynes had (additional) gas-phase formation routes that could compete and even enhance their abundances in the post-shock gas.

\subsubsection{FAUST}
\label{sec:FAUST}

The Fifty AU STudy of the chemistry in the disk/envelope system of Solar-like protostars (FAUST; PI: S. Yamamoto) observed 13 low-mass star-forming regions, each with multiple continuum objects where not all were line rich (\citealt{Codella2021}). Their observations are at an angular resolution of  ${\sim}0.25''-0.35''$ and probe small (${\sim}50$\,au) to large scales (${\sim}2000$\,au). Some of their work focused on the smaller molecules to better understand their chemistry and physical environment of the systems, such as the outflow/jet structure and streamers (e.g., \citealt{Okoda2021}; \citealt{Ohashi2022}; \citealt{Chahine2024}; \citealt{DeSimone2024}; \citealt{Podio2024}; \citealt{Oya2025}). Some other works from this collaboration focused on COMs and the connection of their chemistry with the environment (e.g., \citealt{Codella2022}; \citealt{Frediani2025}). \cite{Bianchi2020} detected lines of multiple COMs toward L1551 IRS5 Class I object. They particularly found a methyl formate to ethanol (C$_2$H$_5$OH) ratio that is similar to Class 0 objects. Therefore, they concluded that not much chemical evolution happens going from Class 0 to Class I. Multiple studies found that shocks can affect the COM emission (\citealt{Vastel2022, Vastel2024}) and a few focused on the connection between hot cores and warm carbon-chain chemistry sources (e.g., \citealt{Imai2022}; \citealt{Okoda2023}).

\subsubsection{GUAPOS}
\label{sec:GUAPOS}
The G31.41+0.31 Unbiased ALMA sPectral Observational Survey (GUAPOS; PI: M.\ T.\ Beltr{\'a}n) observed a line-rich hot core outside the Galactic Center using the entire ALMA Band 3 (${\sim}32$\,GHz bandwidth). They have studied various families of molecules in this dataset including isomers of C$_2$H$_4$O$_2$ (\citealt{Mininni2020}), peptide-like bond molecules (\citealt{Colzi2021}), phosphorus(P)- and S-bearing species (\citealt{Fontani2024}), and generally various N- and O-bearing species (\citealt{Mininni2023}; \citealt{LopezGallifa2024}). A conclusion from these works include, based on correlations among various N- and O-bearing molecules in different objects and with comets, early-phase formation of those species likely on icy grains (\citealt{Colzi2021}; \citealt{LopezGallifa2024}). Another interesting finding from this team was chemical differentiation (seen in peak position and variation in line velocities) of a few O-bearing species compared with other species (\citealt{Mininni2023}).      

\subsubsection{PILS}
\label{sec:PILS}

The ALMA Protostellar Interferometric Line Survey (PILS; PI: J.\ K.\ J{\o}rgensen; \citealt{Jorgensen2016}) observed IRAS 16293-2422 (hereafter IRAS16293) with ${\sim}34$\,GHz bandwidth. This survey has yielded numerous significant findings regarding COMs in protostellar systems. However, here I will focus on some of their main highlights. Most importantly, they provided a complete inventory of COM abundances in this system for COMs with known spectroscopy (e.g., \citealt{Calcutt2018}; \citealt{Jorgensen2018}). Given the large bandwidth and detection of multiple transitions per molecule, their column densities often are measured at a high level of accuracy (e.g., \citealt{Jorgensen2016, Jorgensen2018}) and thus have become a benchmark for many other COM studies. Moreover, this survey resulted in multiple first detections either in the ISM or low-mass protostellar systems (e.g., \citealt{Coutens2016}; \citealt{Lykke2017}; \citealt{Calcutt2018_detect}; \citealt{Manigand2019}). The three new molecules in the ISM that were not a minor isotopologue from this survey were CH$_3$Cl (\citealt{Fayolle2017}), HONO (\citealt{Coutens2019}) and HOCHCHCHO (\citealt{Coutens2022}; \citealt{Muller2024}). In addition, motivated by this rich dataset multiple studies enhanced the available rotational spectroscopy for various COMs which sometimes resulted in a detection (\citealt{Ferrer2023}; \citealt{Muller2024}). Although most works using the PILS data focus on IRAS16293B from the multiple system, they also provided a general inventory of COMs toward IRAS16293A (\citealt{Manigand2020}). Moreover, a few studies considered the agreement of various molecules, including S-bearing ones, with the comet 67P (\citealt{Drozdovskaya2018}; \citealt{Drozdovskaya2019}). For many COMs they found that the abundances toward IRAS16293B are remarkably correlated with those of comet 67P, pointing to the fact that at least a portion of the volatile material of comets and planetesimals may be inherited from these protostellar systems (\citealt{Drozdovskaya2019}). 

\subsubsection{PILS-Cygnus}
\label{sec:PILS-Cygnus}
The PILS-Cygnus is a large bandwidth (32\,GHz) SMA survey of ten intermediate-mass to high-mass systems in the Cygnus-X complex (PI: Kristensen). Not all of these objects show emission from COMs other than methanol (\citealt{vanderWalt2023}). However, based on a few detections, \cite{vanderWalt2023} found that the chemistry of those objects are not particularly correlated with their location within the complex and their distance to the OB2 association. Therefore, they concluded that the external environment plays a less significant role compared to local factors in shaping the chemistry. Another focused study on one of the ten systems, N30 MM1, found a chemical differentiation toward two continuum sources in the system where one mainly shows O-bearing species and the other N- and S-bearing species (\citealt{vanderWalt2021}).

\subsubsection{PEACHES and ORANGES}
\label{sec:PEACHES}

The PEACHES (PI: N. Sakai) observed 50 protostellar systems. Two works from this team focus on simpler S-bearing species (\citealt{Artur2023}; \citealt{Zhang2023}). A major work analyzed COMs in this survey, where they detected them in 58\% of objects (\citealt{Yang2021}). They did not find that this detectability depends on continuum brightness temperature, bolometric luminosity, or bolometric temperature. They found that CH$_3$CN and CH$_3$OH normalized by the continuum brightness have a tight correlation. They also found a similar correlation for all other COMs with a larger scatter. They concluded that this likely points to a common COM chemistry in various objects. 

The Orion ALMA New Generation Survey (ORANGES; PI: A. L{\'o}pez-Sepulcre) considered the OMC-2/3 filament and detected 28 sources, including the multiples, in the ALMA continuum (\citealt{Bouvier2021}). In terms of lines, \cite{Bouvier2022} found warm methanol only toward 26\% of the 19 low-mass protostars that they considered. As this is smaller than what was found for PEACHES, they concluded that their considered filament in Orion has less hot cores compared with Perseus. Pending better future statistics, they concluded that this may point to the different chemical nature of objects in the two regions. 

\subsubsection{PRODIGE}
\label{sec:PRODIGE}

The PROtostars and DIsks: Global Evolution survey (PRODIGE; PIs: P. Caselli and Th. Henning) is a NOEMA large program at ${\sim}1.3$\,mm. The program includes both protostellar systems and protoplanetary disks (\citealt{Semenov2024}). As for the protostellar systems, they cover 30 Class 0/I objects in the Perseus molecular cloud. Multiple studies from this program consider streamers that feed the protostellar systems (\citealt{Valdivia2022}; \citealt{Hsieh2023}; \citealt{Gieser2024}). These streamers were found to affect the observed complex morphological structures of COM emission in the SVS13A system (\citealt{Hsieh2024}). One study from this program considered the $^{12}$C/$^{13}$C ratio of CH$_3$CN and CH$_3$OH in the seven protostellar systems toward which $^{13}$CH$_3$OH was detected (\citealt{Busch2025}). They found $^{12}$C/$^{13}$C ratios of ${\sim}4-30$ which were lower from the local ISM value of ${\sim}68$ by a factor of ${\sim}2-20$. Using chemical models they concluded that the lower ratios might have been inherited from their precursor species for which this ratio is set in the prestellar phase. Another possibility for the lower $^{12}$C/$^{13}$C ratios in their study could be underestimation of the major isotopologue column density due to line optical depth effects. Particularly, lines of $^{12}$CH$_3$OH and $^{12}$CH$_3^{12}$CN with $E_{\rm up} < 300$\,K are most likely optically thick. They did however try to mitigate this by correcting for the optical depth issues.

\subsubsection{SOLIS}
\label{sec:SOLIS}
Seeds Of Life In Space (SOLIS; \citealt{Ceccarelli2017}) is a NOEMA large program targeting seven systems (not all are protostellar objects). There are many studies from this program looking at the chemistry using simple and complex molecules (e.g., \citealt{Fontani2017}; \citealt{Spezzano2020}; \citealt{DeSimone2022}; \citealt{Balucani2024}). Here I will only provide a brief highlight of some of their COM findings. For example, \cite{Bianchi2022} studied deuteration of CH$_3$CN toward a Class I system. They found that CH$_2$DCN/CH$_3$CN in their object is similar to those of prestellar cores. They concluded that based on their results CH$_3$CN is likely formed in the gas phase and then frozen onto dust grains in the prestellar phase. A few other studies considered the shocked regions along the outflow to find a clue on formation mechanism of various COMs (\citealt{Codella2017}; \citealt{Codella2020}), where they found evidence for gas-phase formation of formamide. Another study found evidence for release of methanol from the grains in low-velocity shocks (\citealt{Codella2025}).

\subsubsection{Unnamed surveys}

Here I will highlight a few of the recent multi-object studies with interferometric observations that do not particularly have a name. To keep this section manageable I only discuss studies containing at least five objects or extracted spectra. One such works studied five peaks in two high-mass star forming regions and found that NH$_2$CHO and HNCO are co-spatial pointing to their likely connected chemistry (\citealt{Allen2017}). \cite{Ligterink2020} considered amide-like molecules toward various cores of a massive star-forming region and one finding from this study was that variation in physical conditions within a region can affect production of NH$_2$CN (see also \citealt{Bogelund2019}). \cite{Law2021} used high angular resolution observations ($0.14''$) to assess the complex chemistry toward multiple cores in a massive star-forming region by mapping the column density and excitation temperature (see also \citealt{Williams2023}). They found spatial correlations between NH$_2$CHO/HNCO and CH$_3$OCHO/CH$_3$OCH$_3$. Another study used SMA data toward four massive star-forming regions (\citealt{Law2025}). They found interesting chemical differentiation between N-, O-, and S-bearing species across the regions. They also found similar COM abundance ratios among their objects and those of the low- and high-mass systems in the literature, pointing to early formation of COMs, likely in prestellar ices. One study introduced an automated fitting routine to map various parameters, including column density, for a molecule toward multiple cores of a high-mass star-forming region (\citealt{ElAbd2024}). Given the time-consuming nature of these fits, such automated fitting tools will be particularly important after the ALMA Wideband Sensitivity Upgrade (WSU).

As for low-mass protostellar systems, \cite{Bergner2019} considered 5 Class 0 and Class I protostellar systems and found that three out of five show COM emission (also see \citealt{Domenech2019, Domenech2021}). One conclusion of these papers was that the abundance ratios toward Class I systems seem to be similar to Class 0 systems. Another study by \cite{vanGelder2020} considered O-bearing COMs in seven low-mass protostellar systems but only found COM emission in three (see \citealt{Nazari2021} for N-bearing COMs in this sample). They found that column density ratios for many COMs are similar among different objects, likely pointing to their formation under similar conditions in the cold prestellar phase.

\subsubsection{Upcoming COMPASS and NASCENT-stars}
\label{sec:COMPASS}
Despite the many surveys with ALMA, there are not many line-rich low-mass objects with deep enough data that can detect the less abundant COMs and those with weaker lines. The upcoming Complex Organic Molecules in Protostars with ALMA Spectral Surveys (COMPASS; PI: J.\ K.\ J{\o}rgensen) will help with this issue. They will observe 11 line-rich low-mass protostars at high sensitivity and in a ${\sim}33$\,GHz bandwidth with angular resolution of ${\sim}0.3''-0.5''$. Therefore, this survey will provide high-quality column density measurements of COMs with accuracies similar to those for SgrB2(N2), G31.41+0.31, and IRAS16293 (see Sects. \ref{sec:EMoCA}, \ref{sec:GUAPOS}, and \ref{sec:PILS}). 

Another relevant upcoming survey is the NOEMA Astrochemistry of Cygnus-X Protostars (NASCENT-stars; PI: T. Csengeri). They will cover 17 of the most active star-forming regions in the Cygnus-X complex including both low- and high-mass objects (expected a total of $>60$ individuals). This survey also has a large spectral coverage of ${\sim}31$\,GHz at ${\sim}3$\,mm and ${\sim}15.5$\,GHz at ${\sim}1$\,mm and is expected to constrain the column densities of COMs based on detection of many molecular transitions. The results from these two surveys will considerably add to COM measurements in low- and high-mass objects.

\subsection{Solid phase}

In this section I give an overview of the current solid-phase surveys. Table \ref{tab:ices} presents a summary of JWST icy COM surveys on protostellar systems. The icy COM column densities measure the integrated values along the line of sight and thus measure the bulk abundances from the envelope (Fig. \ref{fig:cartoon}). It is worth noting that in comparison to the gas-phase COM measurements and identifications, the icy COM measurements can be more uncertain due to the complexities mentioned in Sect. \ref{sec:solids}. However, this uncertainty is molecule dependent and some COMs show detection of multiple absorption features which can result in more robust identification and column density measurement.

\begin{table*}
\small
\begin{threeparttable}
    \caption{JWST icy surveys$^{*}$}
    \label{tab:ices}
    \centering
    \begin{tabular}{l l l l l l} 
    \toprule
    \toprule 
Survey &     Instrument       & Wavelength      &   No. of         &      Type of              &   Icy COMs considered \\
&               &      [$\mu$m] &  objects              & objects & in the (sub)sample\\
\midrule  
\hyperref[sec:CORINOS]{CORINOS}    &      MIRI-MRS                        &   ${\sim}5-28$                  &     4                       &         Class 0                     &   CH$_3$CHO, C$_2$H$_5$OH, CH$_3$OH \\
\hyperref[sec:IPA]{IPA}      &        MIRI-MRS and NIRSpec-IFU        &   ${\sim}3-28$                  &     5                       &         Class 0/high-mass       &   CH$_3$OH, CH$_3$CN, C$_2$H$_5$CN   \\
\hyperref[sec:JOYS]{JOYS+}     &       MIRI-MRS (NIRSpec-IFU)          &   ${\sim}2-28$                  &     ${\sim}30\, (17)^{\rm a}$                 &         Class 0/I/high-mass     &   CH$_3$OH, CH$_3$CHO, C$_2$H$_5$OH, \\ &&&&&   CH$_3$OCH$_3$, CH$_3$OCHO \\
&&&&& CH$_3$COOH, CH$_3$COCH$_3$,  \\
&&&&&CH$_3$CN, C$_2$H$_5$CN, NH$_2$CHO \\
\bottomrule
\end{tabular}

\begin{tablenotes}
        \item[*]: This table only presents the medium or high spectral resolution data available with various JWST surveys.
        \item[a]: Some are binaries or multiples.
    \end{tablenotes}
\end{threeparttable}
\end{table*}

\subsubsection{CORINOS}
\label{sec:CORINOS}
The COMs ORigin Investigated by the Next-generation Observatory in Space (CORINOS; PI: Y.-L. Yang) JWST program covers four Class 0 systems observed with MIRI-MRS. Two are known to be line-rich in gas-phase millimeter observations and two are line-poor. The current studies on this program mainly focus on gas-phase lines (\citealt{Salyk2024}; \citealt{Okoda2025}). However, it is worth noting that \cite{Yang2022} identified the ice absorption features that could have contributions from the O-bearing COMs in the ${\sim}7-8.6\,\mu$m region. Therefore, the icy COM abundances will likely be measured and analyzed further in future publications. Particularly, it would be interesting to assess whether there is any relation between the icy COM abundances among their millimeter line-rich and line-poor objects.

\subsubsection{IPA and HEFE}
\label{sec:IPA}

The JWST Investigating Protostellar Accretion (IPA; PI: S.\ T.\ Megeath) program took spectroscopy of five protostellar systems with a range of luminosities and masses (\citealt{Federman2024}) with NIRSpec-IFU and MIRI-MRS. Apart from the gas-phase studies with those data (e.g., \citealt{Federman2024}; \citealt{Narang2024}; \citealt{Neufeld2024}; \citealt{Rubinstein2024}), their current ice studies are mainly on simple ices (\citealt{Brunken2024IPA}; \citealt{Slavicinska2024_HDO}; \citealt{Tyagi2024}). The ${\sim}7-8.6\,\mu$m region is yet to be analyzed, but in the NIRSpec range (${\sim}4.5\,\mu$m) they reported the first tentative detections of CH$_3$CN and C$_2$H$_5$CN in three out of the five objects (\citealt{Nazari2024_ices}), where they found a tentative evidence for enhancement of these molecules in the warmer ices. They also found that the icy abundances of CH$_3$CN/OCN$^-$ show values close to one and thus CH$_3$CN may be an important icy nitrogen reservoir. An IPA follow-up, High Angular Resolution observations of Stellar Emergence in Filamentary Environments (HEFE; PI: S.\ T.\ Megeath), is a JWST large program. It will, apart from imaging of many objects, take spectroscopy of 13 Class 0 protostars with NIRSpec-IFU and MIRI-MRS (${\sim}3-28\,\mu$m). This program is expected to provide valuable data for increasing the sample size of icy COM measurements.

\subsubsection{JOYS+}
\label{sec:JOYS}

The JWST Observations of Young protoStars (JOYS+; \citealt{vanGelder2024}; \citealt{vanDishoeck2025}) is a research initiative that combines data from multiple programs (PIs: E. F. van Dishoeck, M.E. Ressler, T. P. Ray, and T. P. Greene). They observed ${\sim}30$ infrared dark clouds, low-, and high-mass protostellar systems all with the MIRI-MRS and a subset (${\sim}17$) with NIRSpec-IFU spectroscopy. Although a large portion of the JOYS+ studies are on the gas-phase data or simple ices (e.g., \citealt{Ray2023}; \citealt{Brunken2024}; \citealt{vangelder2024_SO2}; \citealt{Caratti2024}; \citealt{francis2025joys}), they have reported the first detection of multiple O-bearing COMs in ices. In particular, \cite{Rocha2024} was the first comprehensive study of the ${\sim}7-8.6\,\mu$m region where species such as HCOOH, CH$_3$CHO, C$_2$H$_5$OH, and CH$_3$OCHO showed a statistically robust detection and further proof for the icy origin of COMs. They found that the abundances of icy COMs correlated within a factor of ${\sim}5$ with the COM abundances in comet 67P. They concluded that this is further evidence for a large portion of cometary COMs being inherited from the protostellar systems (Fig. \ref{fig:cartoon}). Another study from this program on the icy COMs is by \cite{Chen2024} who claimed detection of CH$_3$OCH$_3$ and CH$_3$COCH$_3$ in addition to the three molecules already detected by \cite{Rocha2024} in another source. They found that the icy column density ratios of some of these molecules with respect to methanol match well with those in the gas, while some are higher by ${\sim}1-2$ orders of magnitude. They concluded that the former case likely points to the direct inheritance of gas-phase COMs from the ices. The latter case was suggested to be either due to gas-phase reprocessing following sublimation or physical effects related to the COM distribution in the gas and/or ice. As for icy N-bearing COMs in the JOYS+ sample, \cite{Slavicinska2023} presented an upper limit for NH$_2$CHO in one of their objects, L1527, with NH$_2$CHO/H$_2$O ratio of ${\leq}0.5\%$ which was in agreement with the gas-phase data (see Sect. \ref{sec:ice_comp}). Moreover, analysis of complex cyanides in the JOYS+ sample are presented in \cite{Nazari2025}. Similar to the complex cyanide analysis in the IPA sample (see Sect. \ref{sec:IPA}), only tentative detections of CH$_3$CN and C$_2$H$_5$CN are presented in a few objects of the JOYS+ sample. However, this lack of firm detection for N-bearing COMs can simply be due to the lower abundance of nitrogen compared to oxygen in the ISM (\citealt{Wilson1994}) and specifically for CH$_3$CN and C$_2$H$_5$CN could be the unfortunate positioning of their strongest transitions at around 4.5\,$\mu$m where CO rotational-vibrational gas-phase lines are prominent.

\section{Key results of the surveys}
\label{sec:takeaway}

\subsection{Detection statistics}
\label{sec:detect}

Table \ref{tab:gas} presents a summary of the detection statistics of COMs in the gas phase. Around $2700$ low- and high-mass systems are observed with interferometers that consider COMs. The gas-phase surveys find that methanol detection rates are on average ${\sim}40-50\%$ (i.e., the most detected COM). Various reasons have been proposed for this lack of methanol gas-phase emission in around half of the systems. These reasons are either chemical or physical. From the chemical perspective, methanol could be destroyed in some systems during the infall (e.g., \citealt{Drozdovskaya2014}) or through other processes such as photodissociation (e.g., \citealt{McGuire2017}; \citealt{Notsu2021}). From the physical perspective, dust optical depth could block the emission or decrease it via the continuum-over subtraction effect (e.g., \citealt{DeSimone2020}; \citealt{vanGelder2022}; \citealt{Nazari2024_wind}). Another possibility may be the effect of disks. It has been shown that presence of disks can reduce the temperatures (e.g., \citealt{Murillo2022models}) and thus decrease the emission from water and methanol (\citealt{Persson2016}; \citealt{Nazari2022}). Observationally however, there is mixed evidence on how important the effect of disks is in reducing the emission. For example, \cite{Belloche2020} found no correlation between disk sizes and occurrence of COMs in the CALYPSO sample. On the other hand, \cite{vanGelder2022} found that objects that have large disks (${\gtrsim} 50$\,au) show up to two orders of magnitude lower warm methanol mass (measured from ALMA gas-phase lines) than those with no or smaller disks at similar luminosities.

Ice detection statistics may also help test which of the above options are more important in protostellar systems. If the percentage of objects showing COM or CH$_3$OH detection in ices are found to be similar to those of the gas phase, chemical effects might be the more important factor. On the other hand, if in most objects CH$_3$OH is detected in ices, the physical effects might be the more important reason. From the pre-JWST telescopes, CH$_3$OH was detected in ices of ${\sim}50-80\%$ of the objects (\citealt{Keane2001}; \citealt{Gibb2004}; \citealt{Bottinelli2010}) analyzed. However, the reason for the non-detection in some objects might have been sensitivity limit or inclination angle (\citealt{Crapsi2008}). Therefore, to make a robust conclusion on how important physical and chemical scenarios are for gas-phase COM detection, large-sample ice analysis of COMs/CH$_3$OH with JWST is needed. In particular, considering the same line-rich and line-poor objects with ALMA and JWST might be an interesting path forward. 

\subsection{Column density ratios}
\label{sec:column_ratios}
\begin{figure*}
    \centering
    \includegraphics[width=\textwidth]{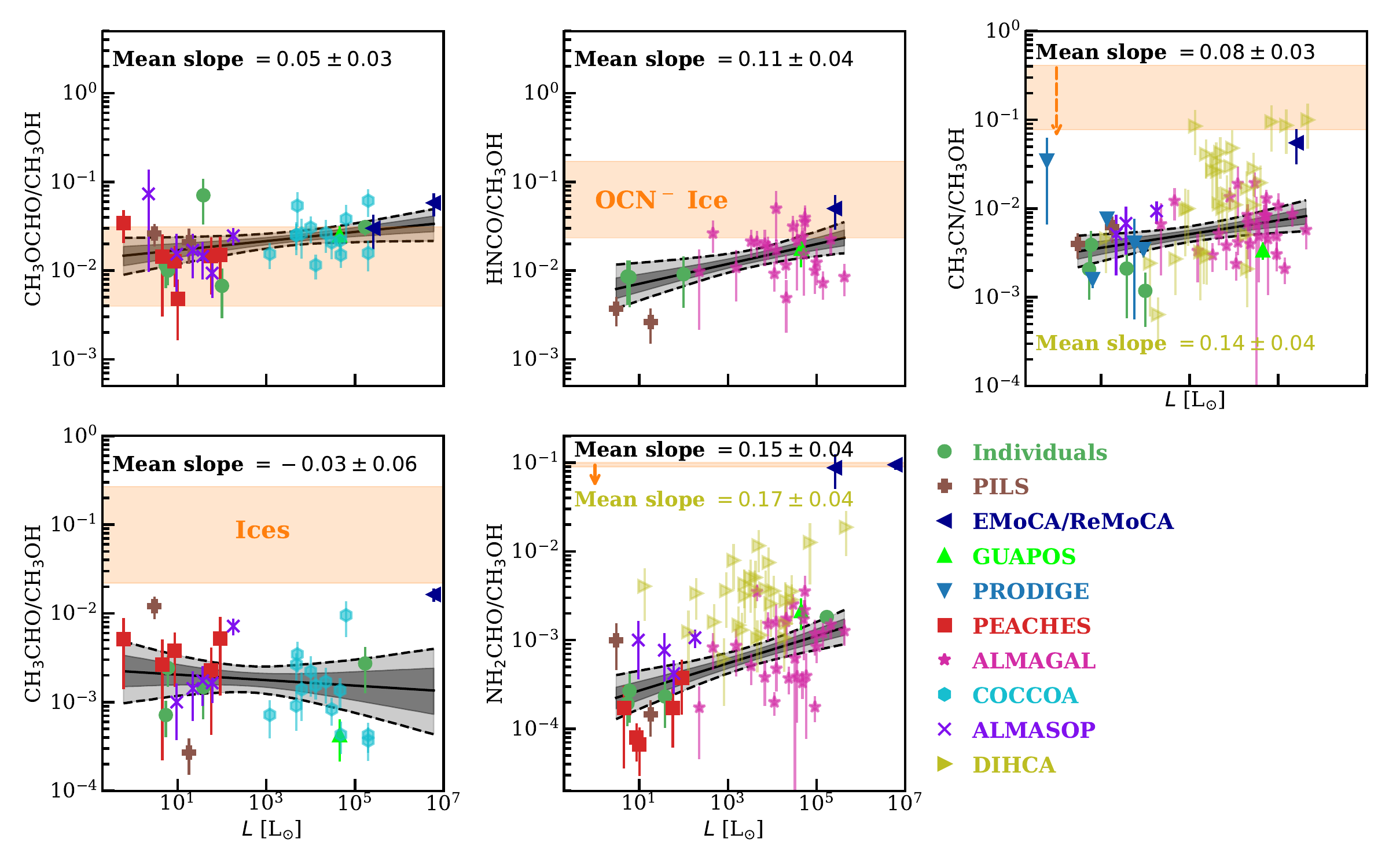}
    \caption{Ratios of various O- and N-bearing COMs with respect to methanol as a function of luminosity. The black solid line is a simple fit obtained using $\chi^2$ method, while the shaded black areas use bootstrapping to find the uncertainty of the fit. The inner darker areas show the 68\% confidence intervals (1$\sigma$) of the bootstrapped predicted y-values at each x-value, while the more extended lighter areas show the 95\% confidence intervals (2$\sigma$). For the NH$_2$CHO and CH$_3$CN panels DIHCA values are not included in the black fit (see text) while the slope is printed in olive color if DIHCA values were included. The PRODIGE object, B1-bS (lowest luminosity), and EMoCA/ReMoCA values, SgrB2(N1-N2), are removed from the black and olive fits in the NH$_2$CHO and CH$_3$CN panels. The highlighted orange regions indicate the range of values of ices for protostellar systems from \cite{Slavicinska2023}, \cite{Chen2024}, \cite{Rocha2024}, and \cite{Nazari2024_ices}. These values include the error bars on the ice measurements and considering the small statistics they are presented as highlighted regions rather than scatter points. The tentative detection of CH$_3$CN in ices is indicated by the dashed arrow and the lowest available upper limit on NH$_2$CHO with a solid arrow. The individual objects and the references for COM measurements are given in Table \ref{tab:refs}.}
    \label{fig:ratios}
\end{figure*}

Figure \ref{fig:ratios} presents a compilation of the literature column density ratios of CH$_3$OCHO, CH$_3$CHO, CH$_3$CN, HNCO, and NH$_2$CHO with respect to CH$_3$OH as a function of luminosity. For the purposes of this section and Sect. \ref{sec:origin}, I consider HNCO among the COMs as it could be chemically linked to other COMs such as NH$_2$CHO (e.g., \citealt{Haupa2019}), is often included in the COM studies, and shows similar distribution to COMs (e.g., \citealt{Lee2022_HH212}). Total luminosity of a cloud is normalized by the continuum fluxes of individual cores, when available, to indicate a zeroth order approximation of luminosities for those individual objects presented in Fig. \ref{fig:ratios}. Figure \ref{fig:ratios} only includes the gas-phase studies that use ALMA data and optically thin lines to find CH$_3$OH, HNCO, and CH$_3$CN. The references for these column densities are given in Table \ref{tab:refs}. Despite the large number of surveys that observe COMs as the main goal or as a side (Sect. \ref{sec:surveys}), the number of surveys suited to analysis of optically thin lines remain limited to high-mass objects where minor isotopologues (i.e., CH$_3^{18}$OH, $^{13}$CH$_3$CN or CH$_3^{13}$CN, and HN$^{13}$CO or HNC$^{18}$O) are more easily detected. 

In Fig. \ref{fig:ratios}, the data points from the PEACHES survey should be taken with caution because the methanol column densities are taken from \cite{vanGelder2022}, while the column densities of the other molecules are taken from \cite{Yang2021}. The two studies use different source sizes (up to ${\sim}0.4-0.5''$) to find the column densities and they either extract their spectra from a region (\citealt{Yang2021}) or from a pixel (\citealt{vanGelder2022}). The PEACHES objects with no detection of minor methanol isotopologues are omitted from the graph. Other data points that also should be considered with care are those from the DIHCA survey. It is important to note that the methanol column densities for those objects were measured in \cite{Sakai2025}, while NH$_2$CHO and CH$_3$CN were measured in \cite{Taniguchi2023} and the two works do not measure these values at exactly the same positions for the same objects. Fig. \ref{fig:ratios} only shows the sources that their right ascension and declination agree within 0.3$''$ which is roughly the beam size of the observations. However, slight differences in the spectra are likely producing some of the scatter in these data points. Moreover, the methanol column densities in \cite{Sakai2025} are measured based on two $^{13}$CH$_3$OH transitions rather than CH$_3^{18}$OH, while in many objects even $^{13}$CH$_3$OH has been found to be optically thick (e.g., \citealt{Chen2024}; \citealt{vanGelder2022}). Therefore, it is possible that for some objects in the DIHCA sample CH$_3$OH values are underestimated. To mitigate potential biases from these data points the black fits do not include DIHCA while the slope printed in olive presents the results of the fits including DIHCA. Although the NH$_2$CHO/CH$_3$OH slope is not significantly affected by including those points, the CH$_3$CN/CH$_3$OH slope is and it likely falls in between ${\sim}0.05$ and ${\sim}0.1$. Nevertheless, the slopes in both cases agree within uncertainties.

Not considering the DIHCA data points, the NH$_2$CHO and CH$_3$CN panels of Fig. \ref{fig:ratios} show the EMoCA/ReMoCA points, SgrB2(N1-N2), as outliers and thus those data points are deleted from the fitted lines to avoid skewing the fit. The reason for SgrB2(N1-N2) being different may be the unique position of SgrB2 close to the Galactic center. However, analysis of SgrB2(N3-N5) with lower continuum brightness than N1 and N2 showed that SgrB2(N2) differs chemically from SgrB2(N3-N5) (\citealt{Bonfand2017}). Therefore, the difference of SgrB2(N1-N2) in Fig. \ref{fig:ratios} might be related to potential differences in the unique physical conditions of SgrB2(N1-N2) and their immediate surroundings that do not include SgrB2(N3-N5). A final data point that is omitted when fitting the data is B1-bS (the lowest in luminosity) from the PRODIGE sample which also is an outlier in panel of CH$_3$CN/CH$_3$OH by a factor of ${\sim}10$. However, this data point has large error bars which agree with the fitted trend. It is worth noting that even though in Fig. \ref{fig:ratios} I have tried to only include objects that use optically thin lines of CH$_3$OH, HNCO, and CH$_3$CN, there are still potential biases to be considered in those measurements. Particularly, the column densities in Fig. \ref{fig:ratios} are mostly found from a limited number of molecular transitions. Therefore, in most cases the excitation temperatures are not well constrained which can affect the column densities. However, this is a much smaller effect than using optically thick lines and normally if excitation temperature is accurate within ${\sim}100$\,K, the column densities should be reliable. Nevertheless, increase in the number of detected transitions will certainly produce more accurate column densities.

Comparing gas-phase ratios of O-bearing COMs with respect to methanol (left two panels) and those containing nitrogen with respect to methanol (middle and right panels), one can see that those containing nitrogen have a shallow positive correlation with luminosity, while the O-bearing COM ratios have no significant correlation with luminosity (also see \citealt{Bhat2023}). This could be related to general differences in formation mechanisms of molecules containing nitrogen and those without nitrogen. On the grains, many O-bearing COMs, including methanol, are generally considered to form from continuous hydrogenation of CO (e.g., \citealt{Fuchs2009}; \citealt{Fedoseev2022}). Therefore, the reason for absence of correlation with luminosity for O-bearing COM ratios could be the chemical link between their formation and methanol formation. For example, chemical models show that one important formation route of CH$_3$OCHO on grains includes CH$_3$O which is also a compound that could be produced as CH$_3$OH forms (\citealt{Simons2020}; \citealt{Garrod2022}). The situation is more complex for CH$_3$CHO as its abundance can also be affected by gas-phase formation routes (e.g., \citealt{Vazart2020}). 

Formation of N-bearing COMs, on the other hand, is less clear. The positive correlation of HNCO, NH$_2$CHO, and CH$_3$CN ratios with luminosity (Fig. \ref{fig:ratios}) may point to a more significant effect of energetic processing and temperature in formation of N-bearing COMs than for O-bearing ones in ices and in the gas. This would be inline with chemical models of \cite{Garrod2022} that find that a significant portion of CH$_3$CN, HNCO, and NH$_2$CHO form in the gas phase at higher temperatures. In addition, \cite{Busch2022} found that N-bearing COMs may have contribution from gas-phase chemical routes toward SgrB2(N1). Moreover, \cite{Nazari2023} (see also \citealt{vantHoff2020_carbon, vantHoff2024}) found that CH$_3$CN may additionally to ices, form in the hot gas closer to the protostars, while they did not find evidence of this hot gas-phase formation for methanol which would support the positive correlation with luminosity in Fig. \ref{fig:ratios}. 

Another reason for the positive correlation with luminosity could simply be the higher binding energies of molecules containing nitrogen. Therefore, for objects with higher luminosities it is easier to get the N-bearing molecules into the gas phase from the grains. Although CH$_3$CN has a similar binding energy to methanol, HNCO may have similar or higher binding energy, and NH$_2$CHO especially has a high binding energy (\citealt{Minissale2022}; \citealt{Ligterink2023}). This order is also aligned with the magnitude of the slopes of their relation with luminosity; CH$_3$CN has the shallowest slope and NH$_2$CHO has the steepest. It is also worth noting that following this argument the reference molecule affects the trends observed. To test this, I also considered the ratios with respect to CH$_3$CN which may have a more similar icy environment and thus sublimation temperature to NH$_2$CHO and HNCO. I found that indeed the slopes of NH$_2$CHO/CH$_3$CN and HNCO/CH$_3$CN decreased for both molecules. A final interesting trend in Fig. \ref{fig:ratios} is the larger scatter of the ratios of CH$_3$CHO/CH$_3$OH and NH$_2$CHO/CH$_3$OH in comparison with the rest at the same luminosity bin. The potential reasons for these differences are already discussed extensively in \cite{Nazari2022_ALMAGAL} and \cite{Chen2023, Chen2024} and they are divided into a chemical or a physical category. Each of these cases are explained in more detail in Sect. \ref{sec:origin}. 

Another important ratio that is considered in multiple studies is the level of deuteration (D/H) in various COMs. The only COM that its deuteration has been measured in a large-enough sample is methanol. Two studies have considered methanol deuteration in a large sample of sources (i.e., $> 20$). One is \cite{vanGelder2022_deuteration} which compiled the literature CH$_2$DOH/CH$_3$OH in low- and high-mass objects and added ${\sim}25$ detection of CH$_2$DOH to the sample using the ALMAGAL data. They particularly found that the higher-mass objects show a lower D/H compared to lower-mass ones. Another study is from the DIHCA team who also added ${\sim}30$ detections of CH$_2$DOH to the sample (\citealt{Sakai2025}). They consistently found lower values of D/H for high-mass objects than low-mass ones. Using chemical models \cite{Sakai2025} concluded that the reason for these lower ratios may be shorter pre-stellar/cold phase timescales in high-mass star-forming regions. On the other hand by comparing with chemical models, \cite{vanGelder2022_deuteration} found that the reason for lower D/H values in higher-mass objects may be either shorter pre-stellar timescales or higher temperatures in the pre-stellar phase for higher-mass objects.

\subsection{Comparison between gas and ice}
\label{sec:ice_comp}

Another clue on formation pathways of COMs can be found by constraining their abundances in the ices and comparing the ice and gas ratios. Figure \ref{fig:ratios} also presents the range of ice abundances measured in the literature (\citealt{Yang2022}; \citealt{Chen2024}; \citealt{Nazari2024_ices}; \citealt{Rocha2024}). HNCO is yet to be found in ices, but its potential progenitor (OCN$^-$) has been detected ubiquitously in protostellar ices since the era of ISO (\citealt{Gibb2004}; \citealt{Oberg2011}) and thus the abundances of OCN$^-$ with JWST data (\citealt{Nazari2024_ices}) are shown on Fig. \ref{fig:ratios}, which agree well with the pre-JWST measurements (\citealt{Boogert2015}). Moreover, because NH$_2$CHO is not yet detected in ices, its smallest upper limit abundance with respect to methanol (assuming a CH$_3$OH/H$_2$O ratio of 5\%) from JWST data of \cite{Slavicinska2023} is presented. Before diving into the details of gas and ice comparison, it is worth noting that the gas-phase and ice measurements trace different scales around the protostar (see Fig. \ref{fig:cartoon}). Gas-phase data trace the warm inner regions while the ice data show the integrated ice abundances along the line of sight through the (disk and) envelope. Moreover, Fig. \ref{fig:ratios} presents ice and gas values for different systems, while the ice-gas comparison might be more useful if done for the same objects. Two sources (B1-c and IRAS 2A) have both gas and ice column densities measured in the literature (also included in Fig. \ref{fig:ratios}) and the direct comparison between their ice and gas has been studied in \cite{Chen2024}.

Comparing the ice and gas-phase abundances in Fig. \ref{fig:ratios}, some ratios agree better and some do not. The cases of CH$_3$CHO and CH$_3$OCHO were discussed in \cite{Chen2024}. Starting with CH$_3$OCHO, the ice and gas ratios agree well. This likely points to formation of CH$_3$OCHO in ices and its smooth sublimation into the gas. Another important factor about CH$_3$OCHO is that it has a similar binding energy to CH$_3$OH at around 6000\,K (\citealt{Mininni2023}; \citealt{Ligterink2023}). This likely results in the two molecules tracing a similar region in the gas and ices which is nicely reflected in the ice and gas agreement in the top left panel of Fig. \ref{fig:ratios}. The situation is more complex for CH$_3$CHO where the ice and gas ratios have around one order of magnitude difference. The CH$_3$CHO/CH$_3$OH ratio is also one that has a large scatter, especially compared to the constant CH$_3$OCHO/CH$_3$OH ratio. The reason for these could be chemical or physical which are further discussed in Sect. \ref{sec:origin}.

\begin{figure*}
    \centering
    \includegraphics[width=\textwidth]{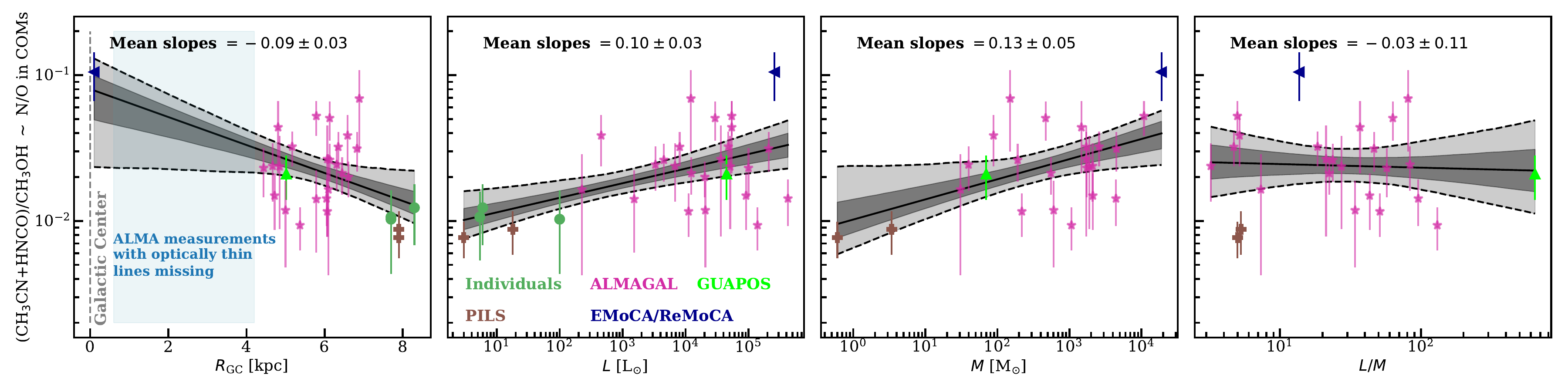}
    \caption{Approximate measure of N/O ratio in COMs as a function of galactic distance, luminosity, envelope mass, and luminosity over mass. The black solid line is a simple fit obtained using $\chi^2$ method, while the shaded areas use bootstrapping to find the uncertainty in the fit. The inner darker and extended lighter areas show the same as Fig. \ref{fig:ratios}. Table \ref{tab:refs} presents the references for these measurements. The relation of N/O as a function of distance is mainly explained by the relation with luminosity and mass, although more deep data with optically thin line for objects in the range of 1-4\,kpc from the Galactic center are needed for confirmation.}
    \label{fig:N_O}
\end{figure*}

Moving to the molecules containing nitrogen, the ice ratio of OCN$^-$/CH$_3$OH agrees better with the high end of gas-phase ratios of HNCO/CH$_3$OH for higher luminosity systems, but is around one order of magnitude higher than the low-end of the gas-phase ratios for lower luminosity systems. This is in agreement with the current knowledge of HNCO formation. From models of \cite{Garrod2022}, HNCO has a major ice formation pathway through reaction of NH and CO. This HNCO can then either desorb from the ices at relatively low temperatures around the protostar (its binding energy is around 1000-2000\,K lower than methanol; \citealt{Ligterink2023}), or turn into OCN$^-$ by reaction with NH$_3$ (e.g., \citealt{Novozamsky2001}; \citealt{Schutte2003}). Another important formation mechanism for HNCO could be through OCN$^-$, which can turn back into HNCO as it desorbs into the gas at higher temperatures closer to the protostar (\citealt{Oberg2009_HNCO}; \citealt{Ligterink2018_HNCO}). The key is that this second reaction occurs at higher temperatures as OCN$^-$NH$_4^+$ thermally decomposes and releases HNCO into the gas (\citealt{Ligterink2018_HNCO}). Going back to the trend seen in the middle top panel of Fig. \ref{fig:ratios}, one can speculate that the observed HNCO for the lower luminosity systems likely resulted from direct desorption of HNCO from the ices at lower temperatures. However, the observed HNCO/CH$_3$OH for the higher luminosity systems may have resulted from the conversion of OCN$^-$NH$_4^+$ to HNCO at higher temperatures which also agrees well with the OCN$^-$/CH$_3$OH ice abundances.

Ratio of NH$_2$CHO/CH$_3$OH upper limits in ices are in agreement with the gas-phase ratios. The tentative ice measurement (indicated by a dashed arrow) of CH$_3$CN/CH$_3$OH ratio is at least one order of magnitude higher than the bulk of the gas-phase measurements. Given the tentative nature of this ice measurement, it is difficult to speculate on its difference with the gas-phase ratios. However, as \citealt{Nazari2024_ices} discussed in detail, this discrepancy could be due to chemical and physical effects which are further explained in Sect. \ref{sec:origin}. Finally, it is important to mention that all panels of Fig. \ref{fig:ratios} show ice ratios that are either equal or higher than the gas-phase ratios. These high ice abundances may point to the general efficiency of ice chemistry for these COMs.

\subsection{N/O in COMs}
\label{sec:N/O}
Table \ref{tab:gas} presents some of the cases where chemical differentiation has been observed across objects or within the same system for the surveys reviewed here. Particularly, a study from the ATOMS survey found that ${\sim}50\%$ of their objects showed variations in peak emission of C$_2$H$_5$CN and CH$_3$CHO (\citealt{Qin2022}). Apart from that, multiple works have reported physical segregation of COMs within the same high- (e.g., \citealt{Peng2013}; \citealt{Gieser2019}; \citealt{Qin2022}; \citealt{Mininni2025}) or low-mass system (e.g., \citealt{Lee2022_HH212}; \citealt{Manigand2020}; \citealt{Frediani2025}). One of the main explanations presented is related to the species binding energy. Molecules with higher binding energies to the ice matrix will desorb at higher temperatures and potentially trace hotter regions. Another explanation for some of the more extended COMs could be non-thermal desorption mechanisms (e.g., \citealt{Busch2022}). Other factors have also been used to explain chemical differentiation within or among objects. For example, source evolution might be a factor (\citealt{Bouscasse2024}; \citealt{vanderWalt2021}) or chemistry along the outflow cavities and shocked regions might affect various molecules (e.g., \citealt{Csengeri2019}; also see Sect. \ref{sec:origin}). Considering the evidence on COM segregation and variations among N- and O-bearing COMs within and in between objects, it is worth evaluating the nitrogen-to-oxygen (N/O) ratio among the current literature sample and its relation with distance from the Galactic center.

Although there are multiple studies on gradient of N/O ratio in the Milky Way and other galaxies using atomic lines (e.g., 
\citealt{Shaver1983}; \citealt{Carigi2005}; \citealt{Rudolph2006}; \citealt{Belfiore2017}; \citealt{Esteban2018}; \citealt{Kumari2018}; \citealt{Maiolino2019}), it is not yet clear whether the same gradient is observed for COMs. An approximation to most of nitrogen in COMs can be found by addition of CH$_3$CN and HNCO column densities, as the two most abundant with respect to methanol in the gas phase. For oxygen, CH$_3$OH is by far the most abundant O-bearing COM with the other commonly observed O-bearing COMs at least having one order of magnitude lower abundances (\citealt{Yang2021}; \citealt{Hsu2022}; \citealt{Chen2023}). Therefore, Fig. \ref{fig:N_O} presents the (CH$_3$CN+HNCO)/CH$_3$OH ratios as an approximation to the N/O in COMs. Although this plot only shows protostellar objects using the measurements found from optically thin lines and data observed by ALMA, adding the value of ${\sim}0.2$ for G+0.693-0.027 (a molecular cloud close to the galactic center) by single dish telescopes (\citealt{Zeng2018}; \citealt{Rodriguez2021}) is consistent with the rest of the data. 

The left panel of Fig. \ref{fig:N_O} shows the COM N/O ratio as a function of distance from the Galactic Center. I note that more data are needed, especially in the $R_{\rm GC} = 1-4$\,kpc regime, for a robust measurement of this relation. Nevertheless, assuming the current gradient, at a first glance, the slope of the fitted line to the data (-0.09) is much larger than those found for our Milky Way from the atomic line studies (\citealt{Carigi2005}; \citealt{Rudolph2006}; \citealt{Esteban2018}; between around -0.04 and around +0.002). This already points to other factors that could be affecting the COM abundances apart from the natural N/O in the galaxy. Considering the relations observed for COM ratios with respect to luminosity in Sect. \ref{sec:column_ratios}, the second panel of Fig. \ref{fig:N_O} shows the N/O as a function of luminosity ($L$). Moreover, the third panel of Fig. \ref{fig:N_O} presents this ratio as a function of envelope mass ($M$). These values, when given for a cloud with multiple cores, are normalized by the continuum fluxes or luminosities, when available, to indicate a zeroth order approximation of envelope mass for individual cores presented. It is interesting that a similarly strong correlation is observed for COM N/O as a function of both luminosity and mass, such that this relation disappears if it is considered with respect to $L/M$ (fourth panel of Fig. \ref{fig:N_O}). This is in practice because the objects that are observed toward the galactic center are also those that are most luminous and massive, likely because of sensitivity limitations. Therefore, for finding the real relationship between N/O and $R_{\rm GC}$ one needs to observe these COMs toward the less luminous and massive objects close to the galactic center.

Nevertheless, it is possible to find the relationship between N/O and $R_{\rm GC}$ by controlling for the dependencies on $L$ and $M$ statistically. Using the \texttt{statsmodels} module in Python (\citealt{seabold2010statsmodels}) and the simple Ordinary Least Squares model, I found the results of the linear regression when fitting $\log_{10}{\rm N/O}$ as a function of $R_{\rm GC}$, $\log_{10}{L}$, and $\log_{10}{M}$ simultaneously. The partial slope for variable $R_{\rm GC}$ was found as $-0.04 \pm 0.04$ with an associated p-value of 0.4 (i.e., not a statistically significant relationship). The 95\% confidence interval for this partial slope was [-0.13, 0.05] which includes the range observed in the literature for N/O from atomic lines. Thus, it is in agreement with those results. It is also interesting to note that the p-values for $\log_{10}{\rm N/O}$ as a function of $\log_{10}{L}$ and $\log_{10}{M}$ in the linear regression are 0.5 and 0.2. Therefore, between the two, the relationship with $M$ is stronger, even though there is still a ${\sim}20\%$ probability of obtaining the same results if $M$ had no effect on N/O.

Finally, the absolute values of COM N/O (ranging between ${\sim}0.01-0.1$) are consistent with the range found in \cite{Rudolph2006} using atomic lines, while only high-end of COM N/O ratios match the low-end from \cite{Esteban2018}. This is in agreement with what was discussed in Sects. \ref{sec:column_ratios} and \ref{sec:ice_comp} about the potentially higher binding energies of N-bearing COMs compared with CH$_3$OH. Therefore, the nitrogen seen in the form of gas-phase N-bearing COMs might not be the entire COM nitrogen budget (ice and gas) for objects with lower luminosities. However, it gets closer to representing the full budget when luminosities and temperatures are high enough to get most of those N-bearing COMs into the gas from the ices.

\section{Chemical or physical origin?}
\label{sec:origin}

A few trends were observed in Sect. \ref{sec:takeaway} that could have either chemical or physical origin. Here, I will consider the reasons behind 1. larger scatter in column density ratios of some molecules and 2. the apparent discrepancy between some ice and gas-phase COM ratios. Starting with the former, the most extreme two molecules that show $> 1$ order of magnitude range in their ratios at the same luminosity bin (Fig. \ref{fig:ratios}) are CH$_3$CHO (\citealt{Chen2023}) and NH$_2$CHO (\citealt{Nazari2022_ALMAGAL}). In addition, from the detected molecules in ices, CH$_3$CHO shows a large difference between the ice and gas ratios. The chemistry of both CH$_3$CHO and NH$_2$CHO is under active debate. For example, \cite{Garrod2022} find that a substantial portion of CH$_3$CHO forms in the gas phase at higher temperatures close to the protostar. Moreover, \cite{Vazart2020} found two gas-phase chemistry routes that could be efficient for CH$_3$CHO using quantum chemistry computations. On the other hand, a few laboratory studies have found efficient ice chemistry routes for this molecule through either hydrogenation of CH$_2$CO (\citealt{Fedoseev2022}) or C$_2$H$_2$ (\citealt{Chuang2020, Chuang2021}). The situation for NH$_2$CHO is similar in that some studies argue for its gas-phase formation (\citealt{Barone2015}; \citealt{Codella2017}; \citealt{Skouteris2017}; \citealt{Lopez2024}), while others suggest that ice chemistry is the more favorable option (\citealt{Jones2011}; \citealt{Dulieu2019}; \citealt{Douglas2022}; \citealt{Garrod2022}). Therefore, it is possible that for some objects in the bottom panels of Fig. \ref{fig:ratios} gas-phase and ice chemical routes both affect the bulk formation and destruction.

However, before concluding on the chemistry of these molecules, physical effects need to be considered. One effect that could bias our conclusions for gas-phase observations is the potential existence of optically thick dust (\citealt{sahu2019implications}; \citealt{DeSimone2020}; \citealt{vanGelder2022}). Another physical factor is the variety in small-scale source structure such as disk size for different objects. Multiple studies have found that considering a disk in a protostellar system, depending on its size can considerably change the temperature (\citealt{Murillo2015, Murillo2022models} \citealt{Persson2016}; \citealt{Jacobsen2018}; \citealt{Nazari2022, Nazari2023Massive}; \citealt{Hsu2023}).

\begin{figure}
    \centering
    \includegraphics[width=0.8\columnwidth]{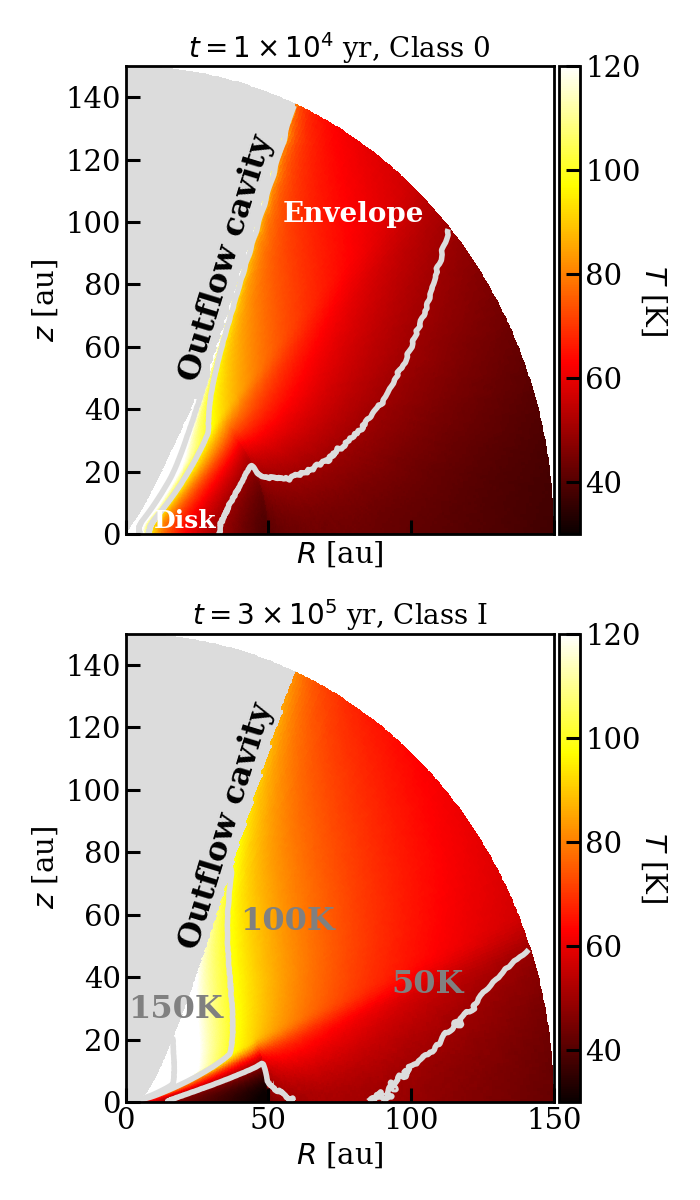}
    \caption{Temperature structure of a Class 0 and a Class I protostellar system. These are taken from \cite{Nazari2024_gap} models with a final stellar mass of 0.5\,M$_{\odot}$ at $t=10^4$\,yr and $t=3\times 10^5$\,yr. The disk radius is assumed as 50\,au for both models, while the disk mass is 0.02\,M$_{\odot}$ and 0.01\,M$_{\odot}$ for the Class 0 and Class I cases, respectively. Other parameters of the model are described in full in Table B.2 of \cite{Nazari2024_gap}.}
    \label{fig:temp}
\end{figure}

Moreover, the temperatures are known to drop for more mature systems (\citealt{vantHoff2020}). This is further shown using radiative transfer models of \cite{Nazari2024_gap} for two typical Class 0 and Class I systems in Fig.\ \ref{fig:temp}. This figure shows the contours of 50\,K, 100\,K, and 150\,K, demonstrating the change in emitting regions for molecules with different sublimation temperatures. Moreover, comparing the top and bottom panels of Fig.\ \ref{fig:temp} highlights that the relation between these temperature contours is not necessarily linear as the system evolves. That is if the emitting area of one molecule is increased by a certain factor for a new object, the emitting area of another molecule with a different sublimation temperature will not necessarily increase by that same factor. 

The data plotted in Fig.\ \ref{fig:ratios} has a large range in $L/M$ and thus maturity of the systems, in addition to a large range in mass and luminosity. This diversity inevitably results in systems with different temperature profiles and emitting regions for various molecules. This phenomenon and its effect has already been extensively modeled and studied in \cite{Nazari2024_scatter}. Where they found that these differences in emitting areas can directly affect the gas-phase column density ratios. This is because in most literature studies the same emitting area is assumed for all molecules toward one object when calculating the column densities due to lack of high-angular resolution data. \cite{Nazari2024_scatter} also found that this bias for two molecules in a ratio can result in as large a difference as one order of magnitude between their ice and gas-phase ratios. 

The two molecules showing a large scatter in Fig. \ref{fig:ratios}, CH$_3$CHO and NH$_2$CHO, have considerably different binding energies (or sublimation temperatures) compared to methanol. The binding energies suggested by \cite{Minissale2022} and \cite{Ligterink2023} for CH$_3$CHO, CH$_3$OH, NH$_2$CHO are 4882\,K, 6621\,K, and 9561\,K in amorphous water ice. Both CH$_3$CHO and NH$_2$CHO have at least 1500\,K difference in their binding energies when compared with methanol, while this difference is much smaller or non-existent when considering the other molecules in Fig. \ref{fig:ratios}. It is also worth noting that the binding energy will be different for the same molecule in two different ice matrices (\citealt{wakelam2017binding}; \citealt{Minissale2022}). Therefore, changes in the ice matrix among various objects will likely magnify the changes in emitting areas among different molecules. Potential paths forward to overcome this issue are discussed further in Sect. \ref{sec:conclusion}. For ices, similar modeling studies are needed to better understand effects of physical factors on ice abundances. 

Observationally, there is mixed evidence for such segregation among molecules as they desorb from the ices (Table \ref{tab:gas} and Sect. \ref{sec:N/O}). For example, in SgrB2(N1) \cite{Busch2022} found evidence for a more extended emission potentially due to non-thermal desorption mechanisms or lower binding energies in the water-poor ice layers and a region above around 100\,K where COMs desorb with water, with no evidence of segregation among them within the 100\,K sublimation radius (Sect. \ref{sec:EMoCA}). On the other hand a FAUST study (Sect. \ref{sec:FAUST}) found evidence for smaller emitting regions of NH$_2$CHO and CH$_3$CHO compared with CH$_3$OH (\citealt{Frediani2025}). Moreover, toward HH212, \cite{Lee2022_HH212} found that NH$_2$CHO and HNCO have smaller emitting sizes than CH$_3$OH.     

Other environmental factors that could affect the COM chemistry are shocks and the amount of UV and cosmic ray radiation. The change in radiation flux has been shown to affect formation and destruction of various COMs in chemical models and laboratory experiments (e.g., \citealt{Oberg2009}; \citealt{Walsh2015}; \citealt{Domenech2020}; \citealt{Garrod2022}; \citealt{delBurgoOlivares2025}). Observationally, there is mixed evidence on how energetic process can affect column densities. For example, an EMoCA study could constrain the cosmic ray ionization rate based on the COM abundances (\citealt{Bonfand2024}). On the other hand, a PILS-Cygnus study found that the external environmental factors such as closeness to the massive stellar cluster (i.e., OB2 association) and potentially higher UV fluence did not have a significant effect on chemistry. The did find however, that local factors related to the outflow or thermal history can have an important effect (\citealt{vanderWalt2023}). 

As for shocks, astrochemical models have found that physical environment of shocked regions along the outflow can enhance production of COMs in the gas phase (e.g., \citealt{Codella2015}; \citealt{Kaminski2017}; \citealt{DeSimone2020_shock}). Other observational studies that specifically target shocks are some of SOLIS where they find evidence for gas-phase formation of COMs in the shocked regions or sputtering from the grains (e.g., \citealt{Codella2015, Codella2020}; \citealt{Favre2020}; \citealt{DeSimone2022}). As for sputtering, it has been shown that the COMs that initially formed in ices could be sputtered into the gas where, according to models, they can survive over the lifespan of the shock (\citealt{Palau2017}). A shock origin for the observed COMs is also claimed toward other objects (e.g., \citealt{Hsieh2024}; \citealt{Hsu2024, Hsu2025}; \citealt{Vastel2022, Vastel2024}). However, it is important to note that such shock origin is not always the case and for various objects a hot core origin where molecules are simply desorbed from the ices is a more likely scenario (e.g., \citealt{Maury2014}; \citealt{Jacobsen2019}). Nevertheless, some of the scatter seen in Fig. \ref{fig:ratios} and differences between ice and gas values could be related to the amount of UV and cosmic ray present in each protostellar system and effect of shocks around some objects.

\section{Final remarks and the path forward}
\label{sec:conclusion}

In this paper, I presented an overview of surveys considering COMs in the gas and ice. The current major takeaways focusing on CH$_3$OH, CH$_3$CN, CH$_3$OCHO, CH$_3$CHO, HNCO, and NH$_2$CHO are summarized below. 

\begin{enumerate}
    \item Transitions of gas-phase COMs have been covered in a total of ${\sim}2700$ objects across multiple surveys. However, not all of those objects show detection of COMs. The most detected COM is CH$_3$OH which on average is detected in ${\sim}40-50\%$ of low- and high-mass systems in the gas phase.
    \item The gas-phase column density ratios with respect to CH$_3$OH for some COMs are remarkably constant in the same luminosity bin (within 1 order of magnitude) across different objects, while those abundance ratios for some other COMs vary more ($> 1$ order of magnitude, Fig. \ref{fig:ratios}). These trends change depending on the choice of reference molecule (e.g., CH$_3$OH vs. CH$_3$CN) however, the following general conclusions regarding the effect of physics and chemistry on the observed values remain valid. 
    \item The gas-phase O-bearing COM ratios with methanol are constant as a function of luminosity, while the N-bearing COM ratios with methanol are positively correlated with luminosity with various degrees. The steepness of that correlation is proportional to icy COM binding energies compared to methanol.
    \item Surveys show that methanol deuteration (CH$_2$DOH/CH$_3$OH) is lower in higher-mass systems than lower-mass ones. The reason has been suggested to be related to the higher temperatures and/or shorter timescales of pre-stellar phase in high-mass star forming regions.
    \item Gas and ice ratios for some COMs with respect to CH$_3$OH agree, while for some others the ice ratios are higher than the gas.
    \item The agreement between gas and ice abundances, the small scatter in gas-phase column density ratios (i.e., tight correlation between molecules), and in general high abundances of icy COMs with respect to methanol point to the importance of ice chemistry in formation of COMs.
    \item The large scatter in gas-phase column density ratios (i.e., anti correlation between molecules) and the disagreement between ice and gas ratios may either be due to chemical or physical factors related to unique source properties such as luminosity, mass, dust optical depth, small-scale structures, radiation field, and existence of shocks. 
    \item Molecules with different binding energies will likely trace different regions around the protostar (Fig. \ref{fig:temp}). Those regions would be different per object, given that each source has a unique physical environment. Therefore, assuming the same emitting size for all molecules in a protostellar system will likely produce a bias in gas-phase column density measurements. This can, in turn, produce the scatter in the gas-phase observations and discrepancy between ice and gas ratios.
    \item The N/O ratios in COMs seem to be positively correlated with $L$ and $M$, while negatively correlated with $R_{\rm GC}$ (Fig. \ref{fig:N_O}). The correlation disappears when the relation is considered with $L/M$. Controlling for the correlation with $L$ and $M$, the slope of the COM N/O relation with $R_{\rm GC}$ becomes similar to that measured from atomic lines. However, more data with optically thin lines, especially in the 1-4\,kpc regime, are needed for confirmation, where the ALMA-QUARKS program (Sect. \ref{sec:ATOMS}) might help. 
\end{enumerate}

Regardless of the most recent improvements, multiple questions remain concerning the formation and evolution of COMs, and the importance of physical factors on chemical interpretation. Less abundant gas-phase COMs (particularly isotopologues of molecules with optically thick lines) have not yet been analyzed statistically in low-mass protostellar systems. This will be largely improved by the upcoming COMPASS and NASCENT-stars results (Sect. \ref{sec:COMPASS}). Another obvious path forward for gas-phase COMs, is high-angular resolution observations. There are only a handful of protostellar systems where COMs have been observed at $< 30$\,au spatial resolution (\citealt{Bjerkeli2019}; \citealt{Lee2022_HH212}; \citealt{Lee2023}; \citealt{Nazari2024_wind}; \citealt{Frediani2025}). Without knowing where COMs trace in the protostellar systems and COMs emitting sizes, our column density measurements and their ratios when molecules have significantly different binding energies, will be erroneous. 

As for the methods of COM analysis, two important factors to be aware of are optically thick lines and number of detected transitions. In particular, for the major isotopologues of the most common species such as CH$_3$OH, CH$_3$CN, and HNCO, it might be more accurate to find their column densities by detecting their minor isotopologues and assuming an isotope ratios. This method has a level of uncertainty due to the assumption of the isotope ratio, however, that is a much smaller effect (factor of ${\sim}2$) than using optically thick lines (potentially orders of magnitude error). Even though it is not possible to always detect more than three transitions of each molecule (especially for the less abundant ones), having more transitions will certainly increase the accuracy of the results. The expected ALMA WSU will greatly help with the efficiency to obtain many more transitions of COMs at higher angular resolutions, thus having the potential to revolutionize COM analysis.

Although gas-phase COM studies have been conducted for decades, research reporting icy COM detections (except methanol) remains in its infancy. Particularly, the ice column densities need to be measured in a much larger sample than currently in the literature. In terms of ice measurements, there are various points of uncertainty that needs to be addressed (Sect. \ref{sec:solids}). Many tools are already available to overcome those issues (\citealt{omnifit}; \citealt{Rocha2021}), but there is a potential to develop further new methods that could help better estimate and improve some of the measurements. More laboratory ice spectroscopy for COMs in a wide range of ice mixtures will also be beneficial. Finally, radiative transfer modeling of the icy envelopes (\citealt{Crapsi2008}), as done frequently for the gas-phase and recently for JWST ices in disks (\citealt{Sturm2023}; \citealt{Bergner2024}), will both improve the ice column density measurements and can illuminate our understanding of effect of physical factors on the ice observations. For example, molecules with different sublimation temperatures will also trace different regions in the envelope which might affect the ice absorption features.

Finally, in the immediate future it is best to take advantage of the synergy between ALMA and JWST. This is because JWST has opened up a new dimension into the COM studies by providing icy COM measurements for the first time. Therefore, analyzing COMs in the gas and ices of the same objects for the largest samples possible before JWST's decommissioning could be a priority. To conclude, the COM formation and evolution has come a long way since the first of these molecules were detected in the ISM (\citealt{Ball1970}; \citealt{Cummins1986}; \citealt{Blake1987}; \citealt{Sutton1991}; \citealt{Helmich1994}; \citealt{Dishoeck1995}). However, many questions still remain open to tackle.


\vspace{5pt}
\noindent \textbf{Acknowledgments}

I thank the two anonymous reviewers for their constructive comments on the manuscript. I am also grateful to Arnaud Belloche, Audrey Coutens, Silvia Spezzano, Tom Megeath, Ewine van Dishoeck, Claudio Codella, Linda Podio, Fr{\'e}d{\'e}rique Motte, Tie Liu, Yuan Chen, Suchitra Narayanan, and Mathilde Bouvier for their insightful comments. The writing of this work was supported by the ESO and IAU Gruber Foundation Fellowship programs. 

\appendix

\section{Column density references}
Table \ref{tab:refs} presents the references for the column densities used in Figs. \ref{fig:ratios} and \ref{fig:N_O}.

\begin{table*}
\small
    \caption{The references for column densities used in Figs. \ref{fig:ratios} and \ref{fig:N_O}.}
    \label{tab:refs}
    \centering
    \begin{tabular}{l l} 
    \toprule
    \toprule 
Source & References \\
\midrule  
SMM1-a & \cite{Ligterink2021} \\
AFGL 4176 & \cite{Bogelund2019AFGL} \\
S68N & \cite{vanGelder2020} and \cite{Nazari2021}  \\
B1-c & \cite{vanGelder2020} and \cite{Nazari2021}  \\
HOPS 108 & \cite{Chahine2022} \\
EMoCA/ReMoCA & \cite{Belloche2016}, \cite{Muller2016}, \cite{Belloche2017}, and \cite{Busch2022}\\
PILS & \cite{Ligterink2017}, \cite{Calcutt2018}, \cite{Manigand2020}, \cite{Coutens2016},\\
 & \cite{Ligterink2017}, \cite{Calcutt2018}, and \cite{Jorgensen2018} \\
GUAPOS & \cite{Colzi2021}, \cite{Mininni2020}, and \cite{Mininni2023}\\
PEACHES  & \cite{Yang2021}, with methanol from \cite{vanGelder2022}\\
ALMAGAL  & \cite{Nazari2022_ALMAGAL} \\
CoCCoA   & \cite{Chen2023}\\   
ALMASOP  & \cite{Hsu2022} \\
PRODIGE & \cite{Busch2025}\\
DIHCA & \cite{Taniguchi2023} and \cite{Sakai2025}\\
\bottomrule
\end{tabular}
\end{table*}

\bibliographystyle{elsarticle-harv}
\bibliography{review}

\end{document}